\documentclass{jfm}
\usepackage{amsmath}
\usepackage{amsfonts}
\usepackage{graphicx}
\usepackage{epstopdf}
\usepackage{epsfig}
\usepackage{hyperref}
\usepackage[noabbrev]{cleveref}
\usepackage{siunitx}
\usepackage{natbib}
\usepackage{xcolor}
\usepackage[utf8]{inputenc}
\usepackage[T1]{fontenc}
\usepackage{lmodern}

\definecolor{carolinablue}{rgb}{0.6, 0.73, 0.89}
\definecolor{pastelblue}{rgb}{0.68, 0.78, 0.81}
\definecolor{pastelorange}{rgb}{1.0, 0.7, 0.28}
\definecolor{cadmiumorange}{rgb}{0.93, 0.53, 0.18}

\DeclareMathOperator{\tr}{tr}

\DeclareMathOperator{\sDiv}{div}

\DeclareMathOperator{\sGrad}{grad}

\newcommand{\cauchystress}{\boldsymbol{\sigma}}
\newcommand{\velgrad}{\mathbf{L}}

\newcommand{\iden}{\mathbf{I}}

\newcommand{\dV}{\ \mathrm{dV}}
\newcommand{\dS}{\ \mathrm{dS}}

\title{Modeling silo clogging with nonlocal granular rheology}

\author{Sachith Dunatunga\aff{1} \and Ken Kamrin\aff{1}\corresp{\email{kkamrin@mit.edu}}}
\affiliation{\aff{1}MIT Department of Mechanical Engineering, 77 Massachusetts Ave, Cambridge, MA, 02139}

\begin{document}

\maketitle

\begin{abstract}
Granular flow in a silo demonstrates multiple nonlocal rheological phenomena due to the finite size of grains. We solve the Nonlocal Granular Fluidity (NGF) continuum model in quasi-2D silo geometries and evaluate its ability to predict these nonlocal effects, including flow spreading and, importantly, clogging (arrest) when the opening is small enough.  The model is augmented to include a free-separation criterion and is implemented numerically with an extension of the trans-phase granular flow solver described in  \citet{dunatunga15}, to produce full-field solutions. The implementation is validated against analytical results of the model in the inclined chute geometry, such as the solution for the $H_{\mathrm{stop}}$ curve for size-dependent flow arrest, and the  velocity profile as a function of layer height. We then implement the model in the silo geometry and vary the apparent grain size.  The model predicts a jamming criterion when the opening competes with the scale of the mean grain size, which agrees with previous experimental studies, marking the first time to our knowledge that silo jamming has been achieved with a continuum model. For larger openings, the flow within the silo obtains a diffusive characteristic whose spread depends on the model's nonlocal amplitude and the mean grain size. The numerical tests are controlled for grid effects and a comparison study of coarse vs refined numerical simulations shows agreement in the pressure field, the shape of the arch in a jammed silo configuration, and the velocity field in a flowing configuration.

\end{abstract}

\section{Introduction}
The behavior of granular media in a silo is a ubiquitous problem arising in various industrial applications \citep{degennes99}.  When a silo opening is large compared to the grain size, the material flows out continuously and the quasi-steady movement inside the silo appears to be relatively well-captured by local constitutive theories, particularly the $\mu(I)$-rheology, which relates the ratio of shear stress and pressure, $\mu$,  to the dimensionless shear rate, $I$ \citep{jop06,staron12,kamrin10,dunatunga15}.  However, contrary to local constitutive theories, which lack an intrinsic size scale, it is well-known that silos can arrest even when the opening is still a finite size \citep{beverloo61,zuriguel03,choi05,sheldon10,thomas13}.  This behavior occurs when the opening size is a small multiple of the grain size, and is characterized by the formation of an arch-shaped bridge of grains that spans the opening.  While the phenomenon of silo clogging can be modeled well using grain-by-grain Discrete Element Method (DEM) simulations (e.g. \cite{martin2009jamming,hidalgo13}), this approach becomes computationally expensive for large systems.  Upscaled relations that have been proposed to capture silo arrest include the well-known Beverloo correlation, an empirical formula that models the outflow rate from a silo directly in terms of the difference between the opening size and a multiple of the grain size \citep{beverloo61}. A blockage criterion is obtained by setting the Beverloo outflow to zero. Another theory for silo clogging reconciles the effect as due to the opening being too small for a finite-sized `spot' of localized motion to pass through \citep{Rycroft2006,bazant2006spot}.  While hybridized DEM-continuum simulations can produce flow stoppage as long as the material near the opening is treated with DEM \citep{yue06}, to our knowledge, no fully continuum constitutive model has been shown capable of capturing the stoppage behavior of silos.  

It is the goal of the current work to present such an analysis. The particular model we shall consider is the Nonlocal Granular Fluidity (NGF) model, which has been shown to capture a variety of nonlocal effects in granular media \citep{kamrin12,henann14,kamrin2019non,kamrin2020quantitative}. Importantly, NGF has been shown to accurately predict size-dependent flow thresholds in a range of quasi-1D geometries \citep{kamrin15,liu2018size}, which encourages its usage for predicting opening-size dependence in silo arrest. As long as our study is restricted to simple granular media (i.e. spherical particles), we also know a priori the model calibration, which enables a more stringent test.  The methodology shown herein extends on the initial work on continuum silo modeling using NGF in the author's thesis \citep{dunatunga17}.

It is worth noting that quantifying silo clogging (and flow thresholds in general) is nontrivial.  As the opening size is reduced, flow intermittency occurs, which can cause one silo realization to clog even though other realizations may continue to flow \citep{sheldon10,thomas13,thomas16}.  Many studies have been conducted to quantify the statistics and scalings of particle release events in the intermittent range \citep{to01,zuriguel05}. Interestingly, it was shown in \citet{thomas13} that while a silo may first clog at one opening size, the critical opening size at which all silos clog is often much smaller and corresponds well with the apparent flow cutoff obtained from extrapolating the Beverloo correlation.  In comparing to a continuum model, our belief is that the continuum solution should predict when the ensemble-averaged flow rate vanishes, which corresponds to this critical size just described.

\section{Nonlocal Granular Fluidity model}
The Nonlocal Granular Fluidity (NGF) model was originally developed as a means to include a length-scale within an otherwise local rheology for steady granular flow. A local rheology is taken to be a constitutive relation in which the local flow state is directly obtainable from the local stress state (and any locally evolving state fields).  The NGF model was initially proposed as a means to reconcile observed departures from the $\mu(I)$ relation in inhomogeneous steady flow geometries, such as annular shear flows and vertical chute flows.  It was later observed that the dynamic form of the NGF model is also able to determine grain-size dependent flow threshold criteria, i.e. when flow in a geometry ceases due to a certain feature of the geometry becoming a small multiple of the mean grain size.  In this regard, the NGF model quantitatively predicts the threshold for flow stoppage  of spherical grains down inclined chutes \citep{kamrin15} known as the $H_{stop}$ effect \citep{pouliquen99,midi04}.  Similarly, NGF predicts flow arrest in 2D cases of disks under annular shear loading, gravity-driven vertical chute flow, and shear load combined with gravity \citep{liu2018size}.

The dynamic (primitive) form of the NGF model can be expressed as an evolution rule for a phase field, $g(x)$, referred to as the `granular fluidity': 
\begin{equation}
 t_0\dot{g}=A^2d^2\nabla ^2g-\Delta \mu \left( \frac{\mu_s-\mu}{\mu_2-\mu} \right)g -b\sqrt{\frac{\rho_s d^2}{p}}\mu g^2
    \label{g_pde}
\end{equation}
where $d$ is the mean grain size, $\rho_s$ is the solid grain density, $t_0$ is a time-scale, $\mu_s$ and $\mu_2$ are the lower and upper bounds taken by the stress ratio $\tau/p\equiv\mu$ during flow, and $\Delta \mu=\mu_2-\mu_s$. The constitutive relation then calls upon the $g$ field to relate stress and strain-rate via
\begin{equation}\label{gmu}
\dot{\gamma}=g\mu.
\end{equation}
The $A^2d^2\nabla ^2g$ term in the evolution of $g$ implies that the spatial diffusion of the fluidity field is influenced directly by the mean particle size.  Hence, flow at a point is not given solely by stress at that point, but rather is influenced by nearby events through the $g$ field.  At steady state, the nonlocal amplitude $A$ is the only new material parameter affecting the flow solution beyond those of the inertial rheology. If $A$ is set to $0$, the steady solutions of the system reduce identically to the inertial $\mu(I)$ rheology.  $A$ is dimensionless and order one; $A\approx0.5$ for glass beads and $\approx 0.9$ for stiff DEM disks \citep{henann13,kamrin14surffric}.  Provided boundary conditions on the fluidity and the velocity, the above becomes a complete constitutive relation.  The model can be extended directly to 3D by defining $\mu$ and $\dot{\gamma}$ based on tensorial invariants \citep{henann13}. 

\section{Grain-size dependent flow thresholds from nonlocal rheology}
To explain how size-dependent thresholds arise in NGF, it is instructive to consider first a paradigmical case, which will be used later for code validation. In recent years, the size-dependence of the flow threshold for flow down an inclined chute has come to be a famous example of nonlocality in granular flow \citep{pouliquen99}. In brief, while local models would predict a thickness-independent repose angle, many experiments show \citep{midi04} that the angle at which a flowing layer arrests grows as the thickness of the layer reduces, and the critical thickness as a function of tilt is known as $H_{\text{stop}}(\theta)$.  See \cref{fig:hstop-phase-and-flow} for geometry details.  One can show from symmetry and quasi-static balance that in the inclined chute, $\mu=\tan\theta$ and $p=zg\cos\theta$ where $z$ measures depth orthogonal to the free surface. Also, from previous data \citep{silbert03}, we can infer that the boundary conditions for the $g$ field are approximately $\partial_n g = 0$ at $z=0$ (the free surface) and $g(z=H)=0$ (the base).

Unlike local models, the NGF model predicts an $H_{\text{stop}}$ curve for this geometry \citep{kamrin15}, which arises due to Eq \eqref{g_pde} developing a bifurcation that causes the $g=0$ solution to lose stability.  That is, while $g=0$, i.e. no flow, is always a solution to the NGF PDE, it is not always a stable solution.  A linear stability analysis of the PDE in the case of an inclined chute reveals  that the stability of the global $g=0$ solution is lost when 
\begin{equation}\label{stop}
    \frac{H}{d}>\frac{\pi A }{2} \sqrt{\frac{\mu_2-\tan\theta}{\Delta\mu(\tan\theta-\mu_s)}}\equiv H_{\text{stop}}(\theta)\, .\end{equation}
We see the right-hand side defines the $H_{\text{stop}}(\theta)$ function\footnote{{\small Or more precisely $H_{\text{start}}(\theta)$ but these two functions are the same in NGF since it does not include effects of hysteresis.}}.  Experiments with glass beads produce an $H_{\text{stop}}$ curve that is close to the solution above when calibrated to the (known) NGF parameters for glass beads. Outside of tilted chutes, the ability of NGF to capture size-dependent flow thresholds was validated in a number of 2D geometries as well, where, just as in the inclined flow case, an analytical solution for the stoppage criterion was found in each case, which predicted the DEM stoppage data to a quantitative level \citep{liu2018size}. 

Jamming of a silo opening shares many similarities with the size-dependent stoppage effect described above.  Most saliently, it is a phenomenon that occurs when the width of the flow domain --- i.e. the silo opening --- competes on the scale of the grain size.  The fact that the grain size is the key length scale implies any continuum treatment would need to be nonlocal to represent this effect.  As such, we seek to determine if the NGF model is capable of predicting silo jamming.

\section{Modeling details for implementing general flow cases}

The NGF equation shown in \cref{g_pde} is intended for dense granular flows.  To use NGF in the silo configuration, the model must be augmented in order to represent separated material.  This disconnected granular phase occurs in the stream of grains that exits a silo when it is flowing, or the material that falls out underneath the arch that bridges the opening when a silo jams.  Without adding a representation of the disconnected phase, determining if a silo jams for a given opening size would be difficult.  One would have to guess an arch shape first and then check if the configuration stays static by seeing if an initially perturbed $g$ field evolves to approach zero everywhere.  Even if the vanishing $g$ solution were stable, it would be unclear if the chosen arch shape were correct --- would the system have found the same arch naturally upon opening a full silo?  Indeed, the arch that forms when a silo clogs should be an emergent geometric outcome and not an input.

In order to properly simulate the formation, or lack thereof, of a stable arch, we extend the NGF model to permit a disconnected phase using a framework similar to that used in \citet{dunatunga15}, which conjoined a separation rule with the local inertial rheology.  As in that work, we consider for simplicity that the disconnected phase is stress-free and defined by when the density, $\rho$, is less than some critical value, $\rho_c$. The dense phase, which abides by the NGF equation, is approximated to be plastically incompressible and defined by $\rho=\rho_c$.  The density field is obtained by enforcing mass conservation throughout the domain.  

On the interface between dense and disconnected phases, the $g$ field requires a boundary condition.  We enforce the Neumann condition $\partial_n g=0$ on these interfaces, guided by our previous observations and prior usage of the Neumann condition along granular free surfaces \citep{kamrin15,silbert03}.  In the event that a zone of granular media reconsolidates from a previously disconnected state, an uncommon event in a silo flow, we want a physically consistent way to re-initialize the $g$ variable.  This resembles the `Stefan free-boundary problem,' frequently encountered in the modeling of solid/liquid fronts in a melting or freezing material specimen. While the classical Stefan problem involves setting multiple boundary conditions for the temperature on a phase front defined by the temperature itself, in our case the question is slightly different in that we must decide boundary conditions for $g$ on consolidation interfaces, which are themselves defined by  $\rho$.  In the results below, we choose to set $g=0$ in newly reconsolidated material, which is the value of $g$ the steady inertial rheology would assume if a finite stress state is uniformly scaled down to zero.  Another option for reinitializing $g$ was used in \cite{dunatunga17} but we noticed it did not make a significant difference in our reported results when we tried it.

\subsection{Generalizing to an elasto-plastic framework}

To accurately implement a model with the components described --- allowing for the possibility of stopped material with an apparent yield stress, dense flow obeying a nonlocal rheology, and separation/reconsolidation of material --- is a \emph{very} challenging task.  We have chosen to address these challenges by representing the model in an elasto-plastic framework, with details provided next. This approach, which we have used previously for a purely local material model \citep{dunatunga15}, essentially turns our proposed rheology into a complex Maxwell fluid by putting the flow rule in series with an elastic element (a  `spring') whose stiffness vanishes when the material density drops below $\rho_c$.  As long as the spring is stiff in the dense (positive pressure) regime, its effect on the flow is negligible.  However, its presence not only assists in implementing the separation/reconsolidation rule, but allows us to model the onset of a true yield stress without the need for viscous regularization.

\subsection{Summary of equations}
We use the standard notation for continuum mechanics as defined in \citet{gurtin10}.
The trace of a tensor $\mathbf{A}$ is given by $\tr \mathbf{A}$, the transpose by $\mathbf{A}^T$, the inverse by $\mathbf{A}^{-1}$, and the deviator by $\mathbf{A}_0 = \mathbf{A} - \frac{1}{3} (\tr \mathbf{A}) \iden$ in 3D.
Scalars are represented by lowercase text, points and vectors by lowercase bold text, and tensors are represented by uppercase bold text (except the Cauchy stress is $\boldsymbol{\sigma}$).
The dot product over vectors is represented with a single centered dot, and simply multiplies the vectors together componentwise and takes the sum, resulting in a scalar (e.g. $\mathbf{a} \cdot \mathbf{b} = \sum_{i} a_i b_i$).
Tensorial contraction is represented with a colon, and multiplies tensors together componentwise and takes the sum, also resulting in a scalar (e.g. $\mathbf{A} \colon \mathbf{B} =  \sum_{i,j} A_{ij} B_{ij}$).
The spatial gradient and spatial divergence operators are given by $\sGrad$ and $\sDiv$ respectively.

Define the strain-rate tensor $\mathbf{D}=(\nabla\mathbf{v}+\nabla\mathbf{v}^T)/2$ and the spin tensor $\mathbf{W}=(\nabla\mathbf{v}-\nabla\mathbf{v}^T)/2$ from the velocity field $\mathbf{v}$. From the stress tensor, define the equivalent shear stress $\bar{\tau}=\sqrt{\sum_{i,j}\sigma_{0\,  ij}\sigma_{0\, ij}/2}$ and hydrostatic pressure $p=-(1/3)\sum_i\sigma_{ii}$, which are used to define the stress ratio $\mu=\bar{\tau}/p$. Let the specific body force (here from gravity) be denoted $\mathbf{b}$, the plastic part of the strain-rate be denoted $\mathbf{D}^p$ (and elastic part $\mathbf{D}^e\equiv \mathbf{D}-\mathbf{D}^p$), and the shear and bulk elastic moduli be denoted $G$ and $K$.

With these definitions in hand, let $\Omega$ be the material domain, let $\Omega_d$ represent the subdomain where material is dense ($\rho>\rho_c$), and let $\Omega_s$ represent the subdomain where material is separated ($\rho<\rho_c$). The continuum system we solve is summarized as follows:\\ \\
\underline{For all $\mathbf{x}\in\Omega$:}\\
\begin{itemize}
\item \emph{Balance of linear momentum, angular momentum, and mass}:
\begin{align*}
    \sDiv \cauchystress + \rho \mathbf{b} = \rho \dot{\mathbf{v}},\ \ \ \ \ \cauchystress = \cauchystress^T, \ \ \ \ \ \dot{\rho}=-\rho\sDiv \mathbf{v}\, .
\end{align*}

\end{itemize}
\noindent \underline{For all $\mathbf{x}\in\Omega_d$:}\\
\begin{itemize}
\item \emph{Elasticity relation}: 
$$\dot{\boldsymbol{\sigma}}-\mathbf{W}\boldsymbol{\sigma}+\boldsymbol{\sigma}\mathbf{W} =K \tr (\mathbf{D}-\mathbf{D}^p)\,\mathbf{1}+2G\, (\mathbf{D}-\mathbf{D}^p)_0\, .$$
\item \emph{Plastic flow rule}: 
$$ \dot{\bar{\gamma}}^p\equiv \sqrt{\sum_{i,j}2D_{ij}^pD_{ij}^p}=g\mu,\  \ \ \ \mathbf{D}^p=\dot{\bar{\gamma}}^p\frac{\boldsymbol{\sigma}_0}{2\tau}\, .$$
\item \emph{Nonlocal fluidity PDE}:
$$
 t_0\dot{g}=A^2d^2\nabla ^2g-\Delta \mu \left( \frac{\mu_s-\mu}{\mu_2-\mu} \right)g -b\sqrt{\frac{\rho_s d^2}{p}}\mu g^2\, .$$
\end{itemize}
\noindent \underline{For all $\mathbf{x}\in\Omega_s$:}\\
\begin{itemize}
\item \emph{Open material condition}:   $$\boldsymbol{\sigma}=\mathbf{0}\, .$$
\end{itemize}

 The above equations are solved simultaneously with the aforementioned (re)consolidation rule for material passing through $\rho=\rho_c$ and appropriate boundary conditions for $g$, tractions, and velocities on walls and free surfaces. Note that although the mathematical model is fully three dimensional, our simulations employ 2D plane-strain assumptions (that is, $D_{zz} = 0$, but $\sigma_{zz} \ne 0$) in our implementation to reduce unnecessary computations.

\section{Numerical Solution: Material Point Method}

\subsection{Introduction}
The Material Point Method (MPM), developed in \citet{sulsky94}, sits at the intersection of Lagrangian and
Eulerian methods. In MPM, the state of a body is tracked in Lagrangian markers
termed Material Points. These material points transfer their state to an
ephemeral background mesh, which is then used to solve the continuum equations
of motion. The deformation of this background mesh is used to inform each of the
material points of their deformation, which is then used to perform the
constitutive update on each material point and move the points, completing the step. At this point,
the ephemeral mesh is no longer needed and can be reset. The key is that the
state of the body is stored entirely on material points, but the solution for
the equations of motion are done on a temporary mesh that resets itself each step. Thus, large deformations
do not accumulate in the mesh, only on material points, and the material points
maintain a consistent and feasible stress state given their individual
deformation histories.

Due to these features, MPM has been used in a variety of granular materials problems. MPM has been used to model individual grains \citep{bardenhagen00} and determine the stresses at contacts along force chains. Others have used continuum descriptions of granular material (e.g. from soil mechanics) to capture phenomena such as landslides \citep{abe13,bandara15,Mast2013}, silo discharge \citep{wieckowski99,Wieckowski2001,wieckowski03,wieckowski11}, and column collapse \citep{mast15}. Our previous work \citep{dunatunga15,dunatunga16} used MPM in this manner with the local $\mu(I)$ rheology to make quantitative and qualitative predictions for well-known phenomena such as the steady-state velocity profiles in chute flow, aspect-ratio dependent run-out scaling during column collapse, the power-law outflow scaling in silo discharge, and the impact of an elastic intruder into a granular bulk. In this paper, we extend our previous method rather significantly to encompass nonlocal effects, to solve nonlocal granular fluidity problems with MPM.  See \cref{compare} for a comparison of MPM to other methodologies.

\begin{figure}
    \begin{center}
        \includegraphics[width=5.4in]{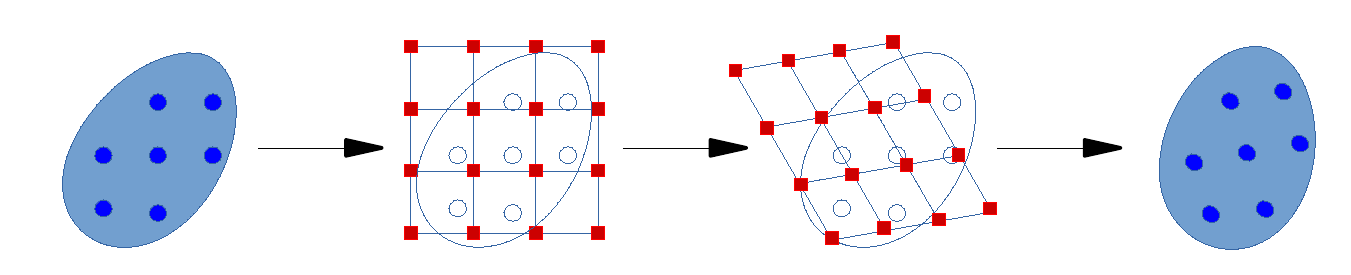}
    \end{center}
    \caption{The procedure to advance an MPM step. First, state information such as mass, volume, and momentum are mapped from the material points to the nodes via shape functions on an ephemeral background mesh (step 1). The mesh exists for a single step to solve the equations of motion (step 2), then the results are projected back to the material points again via the shape functions (step 3). Finally the constitutive update is performed with the new deformation information to update material state variables and the stresses, completing the step. As the background mesh is ephemeral, it can be reset for the next step and cannot accumulate much total deformation.}
    \label{fig:mpm-steps}
\end{figure}

While the theoretical underpinnings of MPM involve a similar construction to
FEM, when writing a computer implementation it may be simpler to conceptualize
the steps as mapping and inverse mapping operators. This is because
material points in MPM are not constrained to a single element throughout the
simulation, while basic FEM typically keeps the carriers of element state collocated with the Gauss points used
during quadrature. For details on how this integration is performed and the variants of MPM which arise from those choices, see \cref{appendix:mpm-integration}.

An overview of the broad steps is presented in \cref{fig:mpm-steps}. As MPM has many variants, the procedure we used to advance an arbitrary time step $t^n$ to $t^{n+1}$
($\Delta t = t^{n+1} - t^n$) is described in words in \cref{appendix:mpm}. These steps can be applied with any constitutive model; we describe the
specialization to the NGF model in further detail below.

\subsection{Constitutive update for the NGF model}

In general, the constitutive update takes deformation-related quantities and
computes the stress state and any state variables. For our case, we are given
the velocity gradient $\velgrad$ over the time step, the density of the
material point $\rho^{n+1}$, the Cauchy stress at the beginning of the time
step $\cauchystress^n$, and the granular fluidity at the beginning of the time
step $g^n$. We wish to calculate the Cauchy stress at the end of the time step
$\cauchystress^{n+1}$ and the granular fluidity at the end of the time step
$g^{n+1}$. 

As implied by the names, at a very high level, the difference between the
nonlocal model and local models is that spatial gradient information is needed
to compute the desired quantities at the end of the step. With the nonlocal
granular fluidity model (and assuming an explicit update), this means we need
to do something akin to the following:

\begin{itemize}
    \item
        Consistently project the value of $g$ on each material point onto a
mesh.

    \item
        Use the mesh to obtain {the discrete Laplacian} 
        of $g$ (call it $\delta^2 g$) at the elemental
level.

    \item
        Consistently project $\delta^2 g$  back to material points.

    \item Use the value of $\delta^2 g$, $g$, and $\cauchystress$ on the
material points to compute the value of $g$ at the end of the step. Note that
now $\delta^2 g$ can be treated computationally as just another state variable,
as it is collocated with the other quantities, though it came from nonlocal information (based on the state of nearby material points).

    \item Use the value of $g$ at the end of the step to determine the
equivalent plastic strain rate $\dot{\bar{\gamma}}^p$ at the end of the step,
which is enough to compute the final stress state.

\end{itemize}

We need to solidify these steps in an actual numerical implementation, as there
are several complications and nuances to implementing the relations and projections.
For pedagogical purposes we do not consider substepping yet, though those
details will be discussed later. A detailed explanation of each step of the
constitutive update we use is provided below.

\begin{enumerate}
    \item
For each material point, check the pressure at the beginning of the time step
$p^n = - \frac{1}{3} \tr \cauchystress^n$ and the density at the end of the
time step $\rho^{n+1}$.  If either of these indicate disconnected material ($p
= 0$ or $\rho \le \rho_c$), the material point does not contribute to the
elemental average of $g$ during this step. Otherwise, the material is dense
during the step, and we need to add its contribution to the background grid. As
we implement volumetric CPDI (see Appendix B), each material point has both a physical volume
and a numerical extent over which it is integrated. If the material point is
entirely contained within an element, we simply add the mass-weighted value of
$g$ to an elemental $g$ accumulator and separately track the elemental mass.
When a material point straddles element boundaries, only the fraction present
in a given element contributes to these quantities (the remainder is added to
the other elements the material point overlaps). After all material points are
accounted for, we divide the mass-weighted $g$ quantities by the total mass in
the element to obtain an estimate for $g$ in that element, which is then used
for our spatial gradient calculations. This splitting of material point extent
was chosen to reduce the cell-crossing error, but we found that simply choosing
to accumulate $g$ to whichever element contained the material point center
produced nearly identical results. We suspect however that covergence of
results may be affected, similar to how uGIMP and MPM convergence rates can
differ.

    \item
Once all dense material points have contributed to the elemental $g$ field
(obtained by dividing the accumulated mass-weighted value by the mass of
contributing material points within that element), we can consider each element
center as a node and use a finite-difference stencil for the Laplacian operator
to calculate
        \begin{align}
            \delta^2 g_{i,j} =
                \frac{1}{(\Delta y)^2} (g^{n}_{i+1,j} - 2 g^n_{i,j} + g^n_{i-1,j}) +
                \frac{1}{(\Delta x)^2} (g^{n}_{i,j+1} - 2 g^n_{i,j} + g^n_{i,j-1}).
            \label{eqn:discrete-laplacian}
        \end{align}
In this case, we use the comma not to indicate differentiation, but simply as
an index based on the logical coordinates of the element ($i$ indicating a row
index and $j$ indicating a column index). In this part of the routine, we
use the value of $g$ at the beginning of the time step when calculating this
quantity, so all terms on the right-hand side are known.

    \item
Now iterate over all material points again; compute the elastic \emph{trial stress,}  $\cauchystress^{tr}$, which is what the updated stress would be if $\mathbf{D}^p=\mathbf{0}$ (i.e. a purely elastic update). While plastic flow can modify the stress deviator from the trial value, in the dense regime the pressure is determined directly from the trial stress since our model assumes no plastic dilation. For any  points with $\tr \cauchystress^{tr}>0$, set the
stress at the end of the step to $\cauchystress^{n+1} = \mathbf{0}$. These
material points are disconnected and not involved in further calculations within this time step,
although they may reconsolidate and participate in later ones.  Note that this check is necessary to catch all open material only for numerical reasons, as the pressure and density may be slightly mismatched since they evolve separately. For all other points, set $p^{n+1}=-\frac{1}{3}\tr\cauchystress^{tr}$. 

    \item
Recall that the evolution equation for $g$ can be written as
\begin{align}
    t_0 \dot{g} = A^2 d^2 \nabla^2 g - \Delta \mu \left( \frac{\mu_s - \mu}{\mu_2 - \mu} \right)g - \frac{\Delta \mu}{I_0} \sqrt{\frac{\rho_s d^2}{p}} \mu g^2.
\end{align}
    We may create a split numerical update for this equation into the following two parts via an intermediate variable $g^{*}$, resulting in the following expressions
\begin{align}
    \frac{g^{*} - g^{n}}{\Delta t} &= \frac{1}{t_0} A^2 d^2 \delta^2 g_{i,j} \\
    \frac{g^{n+1} - g^{*}}{\Delta t} &= \frac{1}{t_0} \left( - \Delta \mu \left( \frac{\mu_s - \mu^{n+1}}{\mu_2 - \mu^{n+1}} \right)g^{n+1} - \frac{\Delta \mu}{I_0} \sqrt{\frac{\rho_s d^2}{p^{n+1}}} \mu^n g^n g^{n+1} \right).\label{gupdate}
\end{align}
    Note that the first expression is easily solved for the value of $g^{*}$, as the discrete Laplacian was obtained from $g$ earlier via \cref{eqn:discrete-laplacian}.

\item
Recalling the plastic flow rule, we can perform an implicit update
by setting $g^{n+1} \mu^{n+1} = (\dot{\bar{\gamma}}^p)^{n+1}$. The plastic shearing $(\dot{\bar{\gamma}}^p)^{n+1}$ is used to correct the elastic trial stress, giving the equivalent shear stress at the
end of the step via
\begin{align}
\bar{\tau}^{n+1} = \frac{\bar{\tau}^{tr} p^{n+1}}{G g^{n+1} \Delta t + p^{n+1}},
\label{eqn:tau_trial}
\end{align}

    \item
After obtaining $g^{*}$, note that \eqref{gupdate} is a semi-implicit
update for the value of $g^{n+1}$. Splitting the $g^2$ term as shown into a
beginning-of-step and end-of-step value allows us to obtain an equation that is
quadratic in $g^{n+1}$. After substitution of $\mu^{n+1}$ obtained from \eqref{eqn:tau_trial},
the quadratic we obtain is

\begin{align}
 0 &= \bar{a} (g^{n+1})^2 + \bar{b} g^{n+1} + \bar{c}\, , \ \ \text{where} \\[6pt]
\bar{a} &= (\mu_2 - \mu_s f) \bar{G} \sqrt{p^{n+1}} + \bar{G} \mu_2 f \bar{\xi}\nonumber \\
\bar{b} &= (S_2 - \bar{\tau}^{\mathrm{tr}} - \bar{G} \mu_2 g^{*} + S_0 f - \bar{\tau}^{\mathrm{tr}} f) \sqrt{p^{n+1}} + (S_2 - \bar{\tau}^{\mathrm{tr}}) f \bar{\xi}\nonumber \\
\bar{c} &= (\bar{\tau}^{\mathrm{tr}} - S_2) g^{*} \sqrt{p^{n+1}}\nonumber
\end{align}

where
\begin{align}
\bar{G} = G \Delta t,\ \ \ S_0 = p^{n+1} \mu_s, \ \ \ S_2 = p^{n+1} \mu_2, \ \ \ f = \Delta t \Delta \mu, \ \ \ \bar{\xi} = \frac{\dot{\bar{\gamma_n}}^p}{I_0} \sqrt{\rho_s d^2}.
\end{align}

Picking the correct root of the quadratic involves some care.
We have the physical constraints that $\mu \leq \mu^{tr}$ and $\mu < \mu_2$. The form of the equation implies that there should be only one solution for $g$ which meets both these constraints, so we solve for both using the familiar expression for the roots of a quadratic equation
\begin{align}
g^{n+1}_{\pm} &= \frac{- \bar{b} \pm \sqrt{\bar{b}^2 - 4 \bar{a} \bar{c}}}{2 \bar{a}}.
\end{align}
As in \citet{dunatunga15}, we may choose to evaluate the roots in mathematically equivalent alternative forms (depending on the signs of $\bar{a}$, $\bar{b}$, and $\bar{c}$) for numerical stability reasons.
We then back-substitute into \cref{eqn:tau_trial} to obtain $\bar{\tau}^{n+1}$ which in turn is used to compute $\mu^{n+1}$ and select the correct root.

\item
    Once the correct $g^{n+1}$ is selected on the material points, use the corresponding equivalent 
shear stress at the end of the step to scale the deviator of the trial stress via
\begin{align}
    \cauchystress^{n+1} = \frac{\bar{\tau}^{n+1}}{\bar{\tau}^{tr}}\cauchystress^{tr}_0 - p^{n+1} \iden
\end{align}
and obtain the stress at the end of the step $\cauchystress^{n+1}$.
\end{enumerate}

The above explanation does not include substepping, which is used to avoid unnecessarily small time step sizes due to small values of $t_0$ compared to the elastic time scales. To implement substepping, we simply use a smaller time-step for the constitutive update alone and repeat the integration until the end of the overall MPM step. The velocity gradient $\velgrad$ is held constant throughout the step, but the stress $\cauchystress$ does update (and as the constitutive update's time-step $\Delta t_{\text{substep}}$ is smaller, the trial stress differs from the case without substepping). The computational savings compared to having an overall smaller $\Delta t$ occur mainly because nodal calculations are elided and because the mapping operations between material points and mesh nodes are reduced to only the initial and final substeps, whereas new mapping matrices must be constructed and used for every full MPM step.

\section{Validation: Inclined Chutes}

\begin{figure}
    \begin{flushleft}
    \begin{tabular}{@{}c@{} @{}c@{}}
        (a) \includegraphics[width=2.4in]{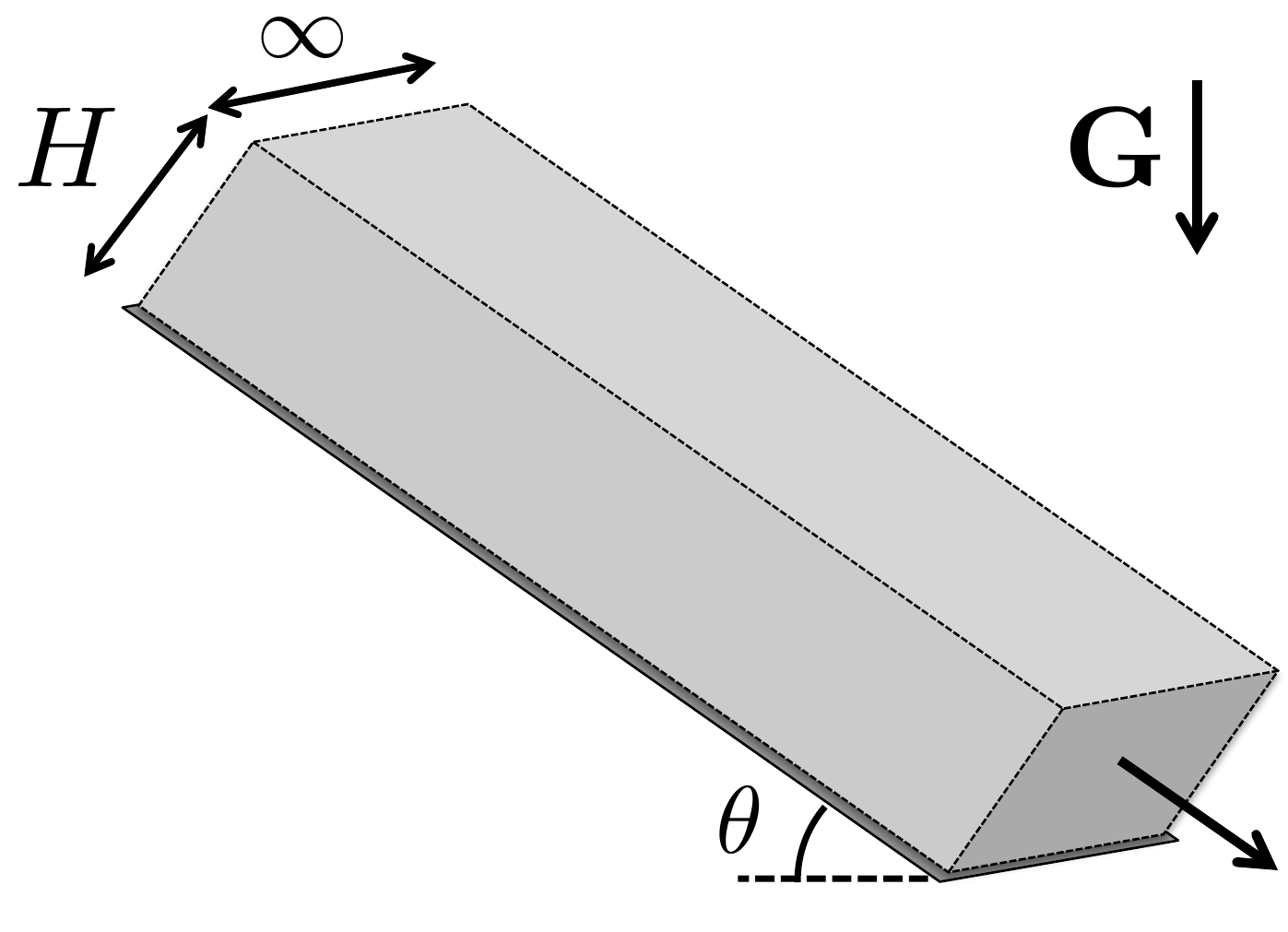} & 
        (b) \includegraphics[width=2.5in]{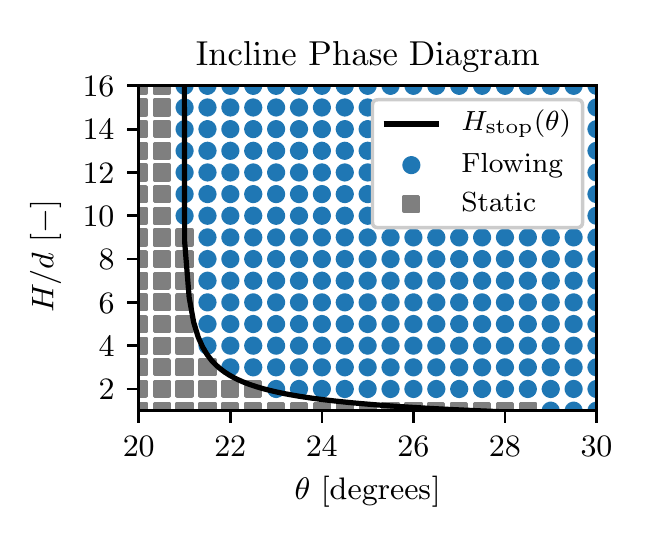} \\
        (c) \includegraphics[width=2.5in]{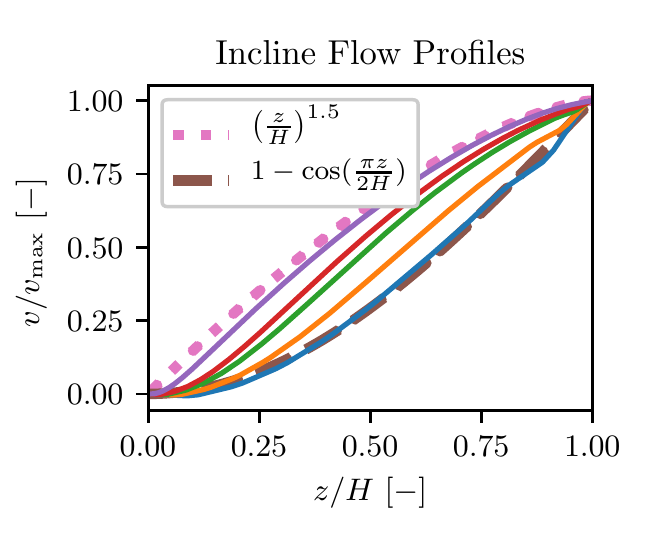} &
        (d) \includegraphics[width=2.4in]{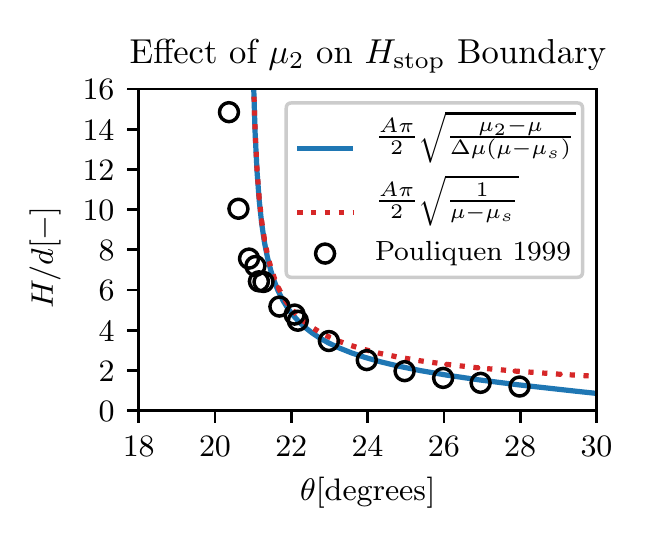} \\
    \end{tabular}
    \end{flushleft}
    
    \caption{Comparing numerical solutions to analytical NGF solutions in the inclined chute.  (a) Configuration of the inclined chute geometry. (b) Phase diagram shows the analytically predicted phase boundary (solid black) and MPM simulation results -- each blue dot is a simulation that appears to flow, while each gray rectangle is a simulation that appears to stop after initial flow. (c) Flow profiles from MPM simulations are taken at an incline of 27 degrees with layer thickness varying through $H/H_{\mathrm{stop}}$ values of 1.9, 2.9, 4.8, 6.8, and 19 (from bottom to top).  (d) Analytical solutions for the fit of the $H_{\mathrm{stop}}$ curve showing the difference between using a common limiting stress ratio  of $\mu_2$ (solid blue) and assuming no limiting value (dotted red). For comparison, experimental data on glass beads is shown (from \citet{pouliquen99}).  Material parameters: (b)-(c) use $A = 0.25$, $\mu_s = 0.3819$, and $\mu_2 = 1.5$, $\rho_s = \SI{2450}{\kilo\gram\per\meter\cubed}$, $d = \SI{5}{\milli\meter}$ and (d) uses inputs for glass beads (from \cite{kamrin15}) $A = 0.48$, $\mu_s = 0.3819$, and $\mu_2 = 0.6435$.}
    \label{fig:hstop-phase-and-flow}
\end{figure}

As an initial validation of our numerical approach, we apply it to the case of wide inclined chutes, cf. Fig \ref{fig:hstop-phase-and-flow}(a).  This geometry is informative because, as previously described, the NGF model has an analytical solution in this case for the thickness-dependent stoppage threshold, which we can use to validate the code's predictions for nonlocally-influenced stoppage.  Moreover, the NGF model has analytical velocity field solutions in the limiting cases of  $H\sim H_{\text{stop}}$ and $H\gg H_{\text{stop}}$, which we can also use to check the code.  

In our simulations, which are implemented as 2D plane strain, we use a single column of elements with $\Delta x = \Delta y = \SI{1}{\milli\meter}$. We represented the body with a uniformly spaced 4$\times$4 arrangement of material points (16 total) per element. Periodic boundary conditions are applied to simulate an infinite chute by using the same node for both the left and right sides, effectively converting this into a one-dimensional problem along the vertical (although material points are still free to move in both dimensions). Recall that for $g$, the bottom boundary uses a Dirichlet $g = 0$ condition while all other sides use a Neumann $\partial_n g = 0$ along the normal to the boundary. We take the grains to be glued to the bottom boundary (via a Dirichlet $\mathbf{u} = 0$ no-slip condition). Common to all included chute simulations, the grain size is given by $d = \SI{5}{\milli\meter}$ and the solid density of grains is $\rho_s = \SI{2450}{\kilo\gram\per\meter\cubed}$. Values for $A$, $\mu_s$, and $\mu_2$ vary as indicated in \cref{fig:hstop-phase-and-flow}.  The specific body force is gravity, $\mathbf{G}$, with $|\mathbf{G}|=9.81$m$/$s$^2$.

Common to all nonlocal simulations, we need to provide initial conditions for $g$, but we require a trick to do this due to a feature of the NGF evolution equation. Note that $g = 0$ is always a solution to the PDE, corresponding to a no-flow state, however it may not always be the \emph{stable} solution. In reality, infinitesimal fluctuations are ever-present and serve to ``seed'' $g$, allowing non-zero $g$ solutions to arise. However, we can exactly represent the zero state during computation, and during a normal update there is no automatic numerical seeding process.

In order to allow a flowing solution to develop at all, we instead use the local $\mu(I)$ model for a short period of time to obtain a flow field, and initialize the value of $g$ based on this local flow.  This nonzero value of $g$ is then evolved to the steady state solution via the NGF equation. Currently, experimental values of $t_0$ are unknown, so for our purposes we choose a small $t_0$ in order to reach steady-state quickly without adversely affecting the time step size. The results are insensitive to details of this choice -- $t_0$ can get an order of magnitude smaller or larger without much issue aside from computational time -- but for completeness, we used $t_0 = \SI{1}{\milli\second}$ with an initialization time of $10 t_0$ through all simulations in this work.  Simulations were run out to $t_f = \SI{20}{\second}$ to get a close approximation to the steady-state behavior, and the time spent in the local state is only a tiny fraction of the total run time.

Each simulation uses a time step size of $\Delta t = \SI{3e-6}{\second}$ and took between 20-30 minutes of wall time on an Intel Silver 4112. Substepping is not used for the presented results (that is $\Delta t_{\mathrm{substep}} = \Delta t$) as the total wall time was low, however we checked that setting $\Delta t_{\mathrm{substep}} = \Delta t / 15$ did not change the outputs.
Figure \ref{fig:hstop-phase-and-flow}(b) shows the analytical solution for $H_{\text{stop}}$ from Eq \ref{stop} together with an array of data points obtained from distinct numerical simulations of our MPM code. Determining if a simulation is truly stopped numerically is difficult, but for our purposes we look simply at the mass-averaged kinetic energy of all the material points in the simulation and compare it to a threshold. We chose \SI{1e-10}{\joule\per\kilo\gram}, but the results are largely insensitive to the specific value. Similarly, other criteria such as average velocity or total momentum result in identical or nearly-identical phase diagrams. It can be clearly seen that our code matches the analytical solution with regard to the placement of the flow/no-flow phase boundary.  Also, when a chute does flow, the NGF model has an analytical velocity solution in certain cases, see \citet{kamrin15} for details.  For shallow layers near $H_{\text{stop}}$, the last term in the NGF PDE \eqref{g_pde} is negligible, leaving a linear system whose solution, upon applying the given boundary conditions, is $v\propto 1-\cos(\pi z/2H)$.  On the other hand, for very tall layers, the nonlocal effect is negligible, and the NGF equations give a velocity field approaching the classical Bagnold solution, $v\propto (z/H)^{3/2}$.  Figure \ref{fig:hstop-phase-and-flow}(c) compares these limiting solutions to output solutions of the MPM code for a wide variety of filling heights.  It can be seen that the MPM solutions capture these limiting cases and indicate how the solution's concavity changes as $H$ increases, a well known effect observed in experiments \citep{midi04,silbert03}.

Regarding flow thresholds, in view of Eq \eqref{stop}, we see that NGF predicts $H_{\text{stop}}$ vanishes when $\tan\theta=\mu_2$; above this tilt angle, the model predicts all chutes flow regardless of filling height \citep{pouliquen99}. This feature is common among many experimental fits for $H_{\text{stop}}$ \citep{midi04}, and leads correspondingly to fits of $\mu(I)$ that also depend on a threshold $\mu_2$ \citep{jop05}. However, there is debate in the experimental literature on the value of $\mu_2$, with some even suggesting it may be infinite \citep{holyoake12}. In fact, a linear $\mu(I)$ fit function is commonly used in practice \citep{dacruz05}, which amounts to setting $\mu_2=\infty$. If one tries to infer $\mu_2$ from $H_{\text{stop}}$ data by observing the tilt at which all chutes flow, complications may arise since this point is not always easy to identify, as it requires measuring $H_{\text{stop}}$ when its value is $\lesssim 1d$.  In fact, even if one chooses $\mu_2=\infty$ in the NGF solution form, the $H_{\text{stop}}$ curve changes only very little.  For example, in \cref{fig:hstop-phase-and-flow}(d) it can be seen that the stopping curves under NGF for a common value of $\mu_2$ and for $\mu_2=\infty$ differ only at the tail and both seem adequate to fit experimental data.  The ambiguity in the choice of $\mu_2$ with respect to inclined flow arrest will come to use in the upcoming section on silo jamming, where certain constraints on the size of $\mu_2$ exist a priori.

\section{Silos}
\subsection{Silo geometry and numerical parameters}

\begin{figure}
    \begin{flushleft}
        (a) \includegraphics[width=2.3in]{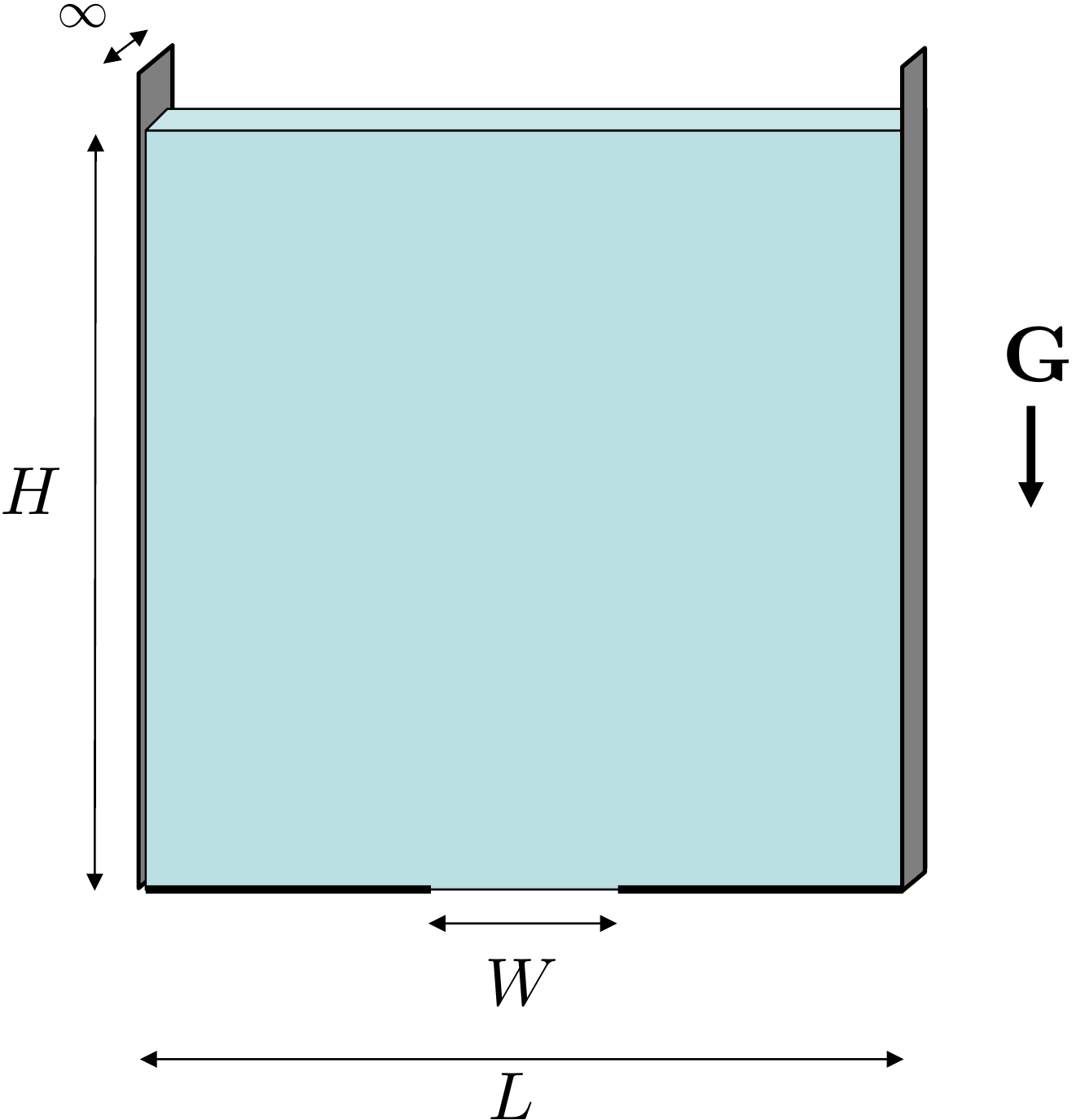} \ \ \ \
        (b) \ \begin{minipage}[b]{2.2in}
                \includegraphics[width=1.8in]{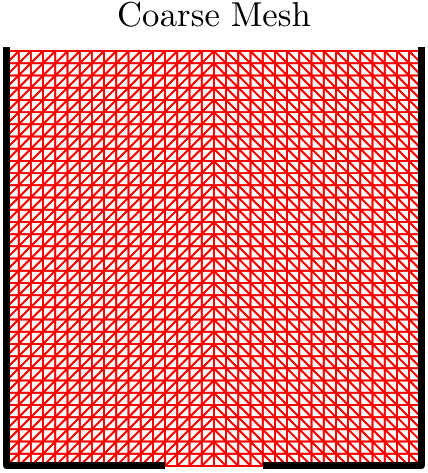}
                \vspace{0.4in}
              \end{minipage}
    \end{flushleft}
    \caption{The geometry of the silo is shown in the left panel (a). We kept the height H, width L, and orifice width W consistent throughout the simulations and changed the material parameters, mainly grain size $d$ and nonlocal amplitude $A$. The background mesh used by the MPM scheme for coarse cases is shown in (b) to aid in understanding the apparent shape of the arch in static cases.}
    \label{fig:silo-geometry-setup}
\end{figure}

Our implementation is in 2D plane strain, corresponding to a thick silo in the out-of-plane direction. When choosing the geometry of the mesh, note that the silo needs to have enough elements across the orifice to initiate and resolve the flow (in a flowing case), but to reduce computation time we would like to keep this number manageable. We chose 8 units across the orifice, with each unit being a \SI{0.00175}{\meter} by \SI{0.00175}{\meter} square. Although the background mesh is ephemeral in MPM, since we solve the equations of motion over this mesh it does have an influence on the space of possible solutions on the material points. As with the inclined chutes, each square unit contained 16 material points (four per spatial dimension). In order to reasonably match the geometry of an arch, each unit consists of two triangular elements with the diagonal going from bottom left to top right for the left half of the silo and vice-versa for the right half (see panel (b) of \cref{fig:silo-geometry-setup}). We show later that results are insensitive to the exact number of elements as long as there are enough to resolve flow features around the orifice. A full convergence study is not attempted in the present work, as the computational time and space required quickly becomes prohibitive due to the explicit parts of the scheme and the number of auxiliary variables respectively.

Material parameters are presented in \cref{tab:common-material-parameters}, which are largely consistent with values commonly used for a 3D packing of glass beads. Of note however is the large value of $\mu_2$, explained as follows. At a jammed arch in a plane-strain silo, the stress state must be compressive along the arch, but vanish normal to it because it is a free surface.  Thus, the stress state at the arch must attain $\mu = 1$; this fact is true regardless of the constitutive model.  To properly capture this stress state, it follows that $\mu_2$,  the feasible upper limit for $\mu$ in models like ours that extend the $\mu(I)$ rheology, cannot be less than $1$.   The constraint is slightly weaker for fully-3D silos with a hole orifice, where $\mu_2$ must exceed $\sqrt{3}/2\approx 0.866$ to admit a jammed state. Both these values are higher than, for example, $\mu_2=0.6453$ reported in \cite{jop05}, but recall from the section on inclined chutes that there is ambiguity on the value of the upper limit of $\mu_2$ and little experimental data in that regime.  We observed in our simulations that above $\mu_2\sim 10$, the solution becomes insensitive to the value of $\mu_2$ so we simply set $\mu_2$ to an asymptotically high limit, giving effectively a linear $\mu(I)$ fit in the local limit. 

\begin{table}
    \centering
    \begin{tabular}{c | c}
         Parameter & Value \\
         \hline
         $E$ & \SI{1}{\mega\pascal} \\
         $\nu$ & 0.45 \\
         $\mu_s$ & 0.3819 \\
         $\mu_2$ & 50.0 \\
         $t_0$ & \SI{1e-3}{\second} \\
         $\rho_s$ & \SI{2450}{\kilo\gram\per\meter\cubed} \\
         $\rho_c$ & \SI{1500}{\kilo\gram\per\meter\cubed} \\
         $b = \Delta \mu / I_0$ & 1.0 \\
    \end{tabular}
    \caption{Common material parameters used for the silo simulations. Our simulations vary $A$, hence it is not present in this table, but instead will be specified on a case-by-case basis.}
    \label{tab:common-material-parameters}
\end{table}

Each simulation is run with the same orifice size $W$, given by \SI{0.014}{\meter} (measured from fixed-node to fixed-node), but the grain diameter $d$ is varied via the material properties to set the ratio $W / d$. The total width of the silo $L$ is fixed at \SI{0.595}{\meter} (34 units), and, importantly, the same background mesh and material point density is used for all simulations so as to control the possible influence of numerical length scales. The initial fill  height $H$ is also fixed at \SI{0.595}{\meter}.
Initially the material point configuration is assigned a lithostatic pressure; the density of each material point is set to be commiserate with this pressure by computing $\rho = \rho_c K / (K - p)$, where $K$ is the bulk modulus and $p$ is the pressure of that material point. The density is initially set by changing the mass of the material point -- all material points have the same initial volume in these simulations. However, as in \citet{dunatunga15}, the true volume of material points evolves over time (though the numerical extent they are integrated over does not, see Appendix B).

Similar to the inclined chutes, we set the bottom of the silo to have Dirichlet $g = 0$ and $\mathbf{u} = \mathbf{0}$ (no-slip) boundary conditions -- note these are only applied along the solid portion of the boundary (i.e. they do \emph{not} apply directly over the orifice, which instead allows the granular material to act as a free surface with $\partial_n g = 0$ and a traction-free condition). The sidewalls allow friction-free sliding (perfect-slip), but do not allow material points to go through the boundary; that is $v_n = 0$ while the transverse direction is unhindered, and $\partial_n g = 0$.

No motion is imparted to the material points as an initial condition. Instead, the retaining wall at the orifice is removed upon the first step of the simulation and flow is allowed to develop ``naturally''.  Recall from the section on inclined chutes that we initialize $g$ by running the local model for a short period of time. As the local model will always produce a flow field regardless of the ratio of orifice to grain size, this procedure will always result in a nonzero $g$ field.
After initializing with the local model, the numerical procedure detailed above for the nonlocal model is used to determine if the solution decays towards a stable $g = 0$ state corresponding to a jammed silo or evolves to obtain some other steady flowing state.

We found the results largely insensitive to the specifics of how the seeding procedure was performed; changing from a local model $\mu(I)$ to a simple rate-independent Drucker-Prager material only slightly affected transient states after switching to the nonlocal procedure, and decreasing or increasing the local model run time similarly had almost no discernible effect on the final state. There were two notable exceptions to this. First, if the material points deviate too much from their initial positions, finding a non-flowing solution became difficult, at least at the spatial and temporal sampling parameters we had used. This is likely because a strong arch of material points needs to develop simultaneously without any holes, and as points move further from their initially ideal placement, numerical noise in MPM makes this unlikely. Refinement would likely mitigate this issue, as the crossing noise can be reduced and the additional material points are more likely to sample a good configuration for numerical purposes, but we did not explore this further. Secondly, if the initialization process were done too quickly, the flow field (and therefore $g$) may not have enough time to establish through the system and ends up too small in magnitude to seed the simulation in light of numerical damping. 

Outside of initialization, elastic waves through the system may serve to jostle a clogged silo away from that state. While some of these effects are physical, real systems also include elastic-wave damping effects not included in our model; moreover, we are most interested in the elastically rigid limit where elastic waves would not occur anyway. Adding bulk viscosity could mitigate this issue but we felt this would interfere too much with the solution.  Instead, we implemented an elastic-wave absorbing boundary by zeroing-out the speed of any material point moving \emph{upward} that is located at y-position \SI{0.49}{\meter} (28 units) or higher.  Upward motion is entirely caused by elastic waves in this problem since gravity points down. This absorbing condition is located sufficiently far from the orifice that we believe it should not affect any true dynamics near the opening when the system clogs or in flowing cases while serving to remove unwanted elastic waves. Results were largely insensitive to the position or strength of the penalty (e.g. moving the dividing line down by \SI{8}{\centi\meter} did not change the results of our analysis, nor did implementing the penalty as a halving of upward velocity instead of zeroing).

\subsection{Silo jamming phase diagram}
 Even if it may be obvious to the eye, determining a precise criterion for when a silo configuration is jammed in a numerical simulation is a nontrivial exercise. It can be comparable to studying avalanche effects and the presence of open/falling material can trick kinetic-energy-based criteria for solidification \citep{aranson01,martin2009jamming}. 
Instead, we use a heuristic to mark states as ``likely static'',
``likely flowing'' and ``unsure''.
\begin{figure}
    \begin{flushleft}
        (a)\includegraphics[width=2.3in]{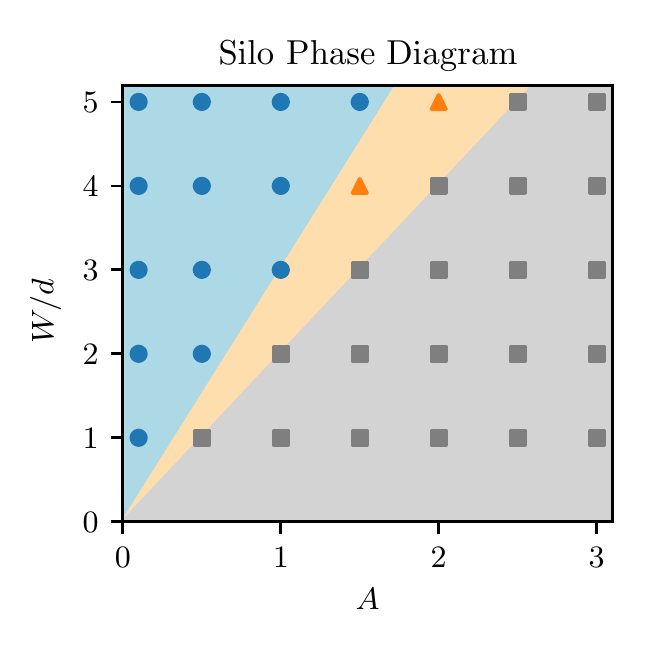} \  
        (b)\ \ \ \includegraphics[width=2.4in]{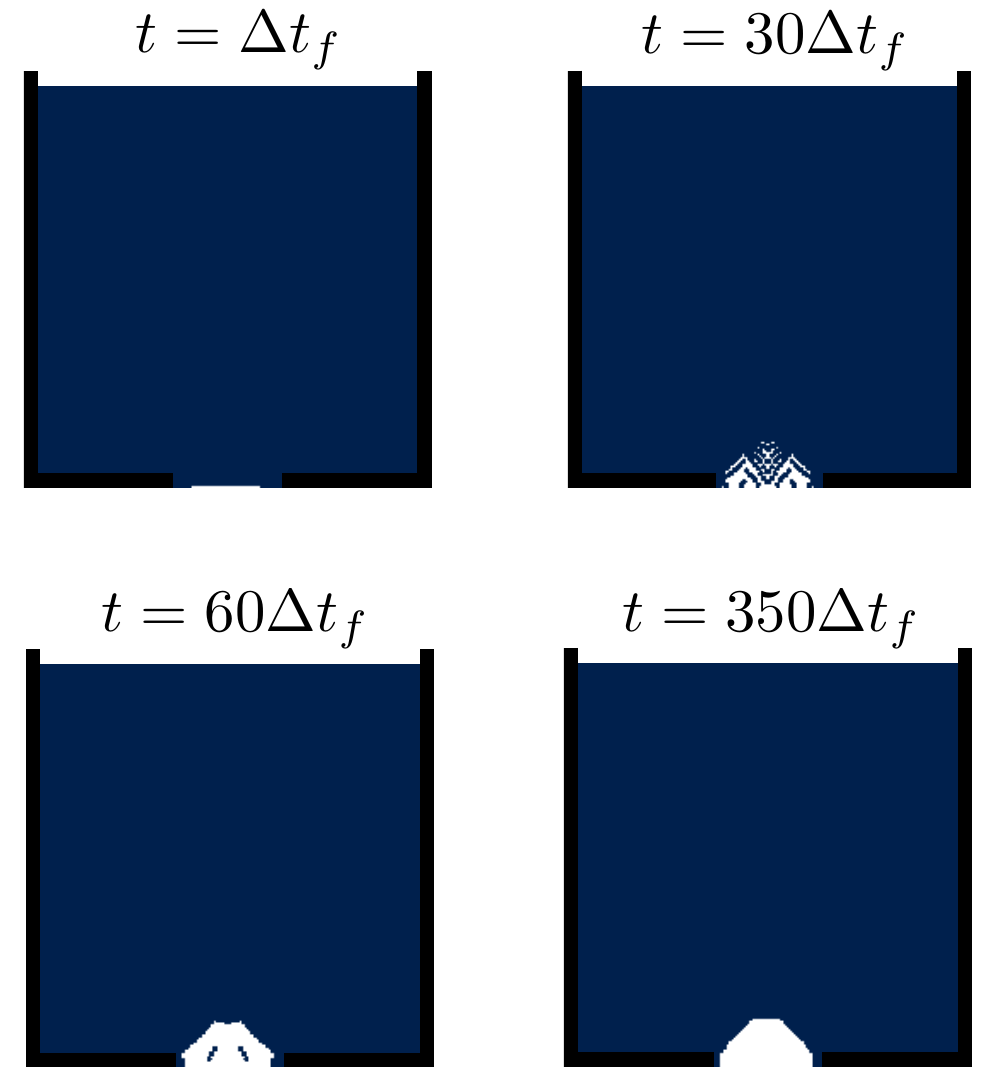} 
    \end{flushleft}
    \caption{ (a) Silo phase diagram varying nonlocal amplitude $A$ and the relative opening size $W/d$. A blue circle means the simulation flows, a gray square means the simulation attains a static configuration, and an orange triangle indicates that we were unsure how to classify the simulation -- see the main text for details on the classification criteria. Note that the geometric properties of the simulation were held constant and that only the indicated material properties (that is, the grain size $d$ and $A$) were varied here. Other material properties were held constant for these tests. Note that the diagram can be partitioned into the three regions with only two lines which go through the origin.  (b) Still frames of the arch formation process for the $A = 1.0$, $W/d = 1.0$ case, which is a static simulation in our classification.}
    \label{fig:silo-phase-diagrams}
\end{figure}
If the average velocity of the \emph{dense} material is small and at most exhibits decaying oscillatory waves, we suspect that the material is in a static
configuration. Additionally, if only a very small number of grains are leaving per unit time,
we can be more sure that the configuration is static. Conversely, if the
average velocity points downward, is large, and does not change size much, we suspect that
the material is flowing; we can be even more sure if a large number of grains are
leaving per unit time. To make this more precise, in our simulations we check if the most recent zero
crossing of the vertical component of velocity happened within the last ten
percent of the simulation; if so, we give a point in favor of a static
configuration. If the last zero crossing happened earlier than halfway through
the simulation, we give a point to the flowing state; in between, we are
unsure. We then check the number of material points leaving between each saved
frame of the simulation; if this is less than or equal to one on average in the
last tenth of the simulation, that is a point in favor of a static
configuration. If this is greater than ten per frame on average, that is a
point in favor of a flowing simulation; in between, we are unsure. Both tests
must agree (i.e. two points are required for us to mark the simulation as
static or flowing) -- any other state is marked as unsure. Visual inspection of the last few frames of a selection of simulations was also done for confirmation.

Simulations took approximately 2 hours on an Intel Silver 4112, with faster-flowing simulations finishing earlier as they had fewer material points over time. In all silo simulations, we used a substepping time of $\Delta t_{\mathrm{substep}} = \Delta t / 150$, which we estimate from smaller cases reduced the wall time by a factor of two.
Using the criteria outlined above, we plotted this data on a lattice with simulations having varying $A$ and $W/d$ ratios (as $W$ is fixed, we simply changed the material grain size $d$ to achieve the desired ratios). The value of $A$ is either $0.1$, $0.5$, $1.0$, $1.5$, $2.0$, $2.5$, or $3.0$ while $W/d$ must be either $1$, $2$, $3$, $4$ or $5$. The results are shown in \cref{fig:silo-phase-diagrams}(a). Note that we see a separation into three regions (flowing, static, and unsure) which can be well-described simply by two lines of different slope; the ratio of $W/d$ to $A$ seems sufficient to predict if a simulation will flow or reach a static state.

This is not too surprising if we consider the form of the NGF PDE and the setup we used. Note that the term $d$ almost always appears with $A$ -- it only shows up alone in the $g^2$ term for describing the local rheology, which we believe only has a minor effect on the stability of the solution. Since $W$ is fixed in these simulations, lines of constant $Ad$ should therefore exhibit very nearly the same stability behavior, as the only difference would appear in the $g^2$ term. In our graph, starting at the origin, we can fan out to form these lines of constant $Ad$, and we do see a physically-reasonable boundary between flowing and non-flowing cases. As a benchmark against experimental data, note that at $A = 0.5$ (an approximate value for {3D beads}) the flowing/static boundary occurs somewhere between $W/d = 1$ and $W/d = 1.5$. This corresponds well with experimentally observed values for the critical $W/d$. In the flat-bottomed quasi-2D silo, which we are attempting to model,  $W/d\approx 1.0$ was obtained for beads in \cite{choi05}.  This is similar to the range of $1.0-1.16$ that was obtained in \cite{Mankoc2007} for beads, albeit for a circular orifice. As before, we note that since we have a continuum solver, we achieve the static configuration when the ensemble average behavior would be to form a static arch. In a single experiment, it may be possible to observe orifice openings of larger size that nevertheless still can find a static configuration, but these become rare as the ratio $W/d$ increases. %
Figure \ref{fig:silo-phase-diagrams}(b) shows an example of an arch forming in one of our continuum simulations (where $\Delta t_f = \SI{1/240}{\second}$) when the silo  opening is sub-critical, with material points below the arch falling out leaving the rest of the silo material intact as the fluidity decays to zero.  The numerical procedure models the outline of the material domain from edges of the background mesh geometry, which leads to a `polygonal' arch shape; as the mesh refines more, a rounder arch shape emerges (see Sec \ref{refine} and Supplemental Movie). 

\begin{figure}
    \begin{center}
    \begin{tabular}{@{}c@{} @{}c@{} @{}c@{}}
        \includegraphics{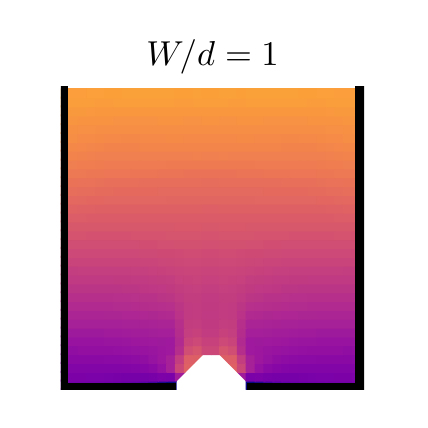} &
        \includegraphics{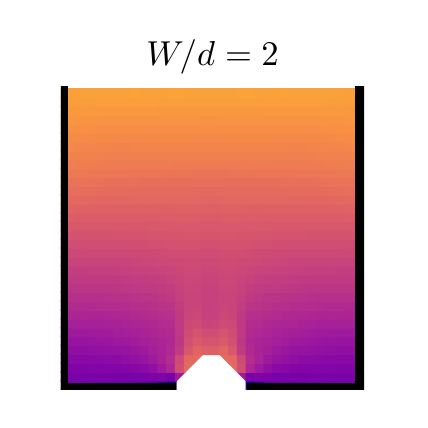} &
        \includegraphics{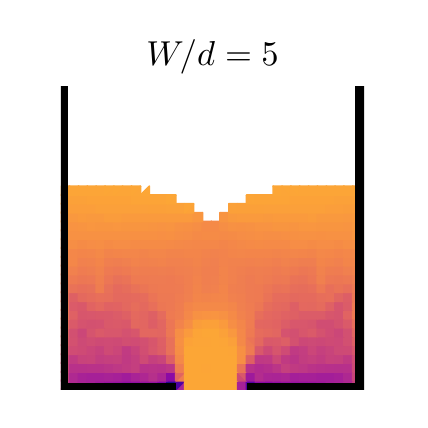} \\
    \end{tabular}
    \end{center}
    \begin{center}
        (a) \includegraphics{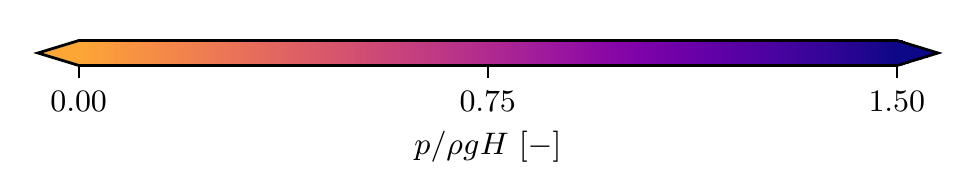}
    \end{center}
    
    \begin{center}
    (b) \includegraphics{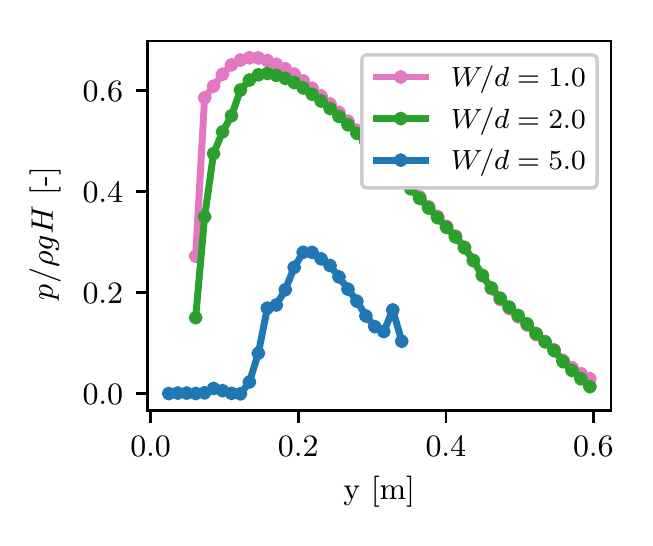}
    \end{center}
    
    \caption{(a) Visualization of pressure in silos with $A = 1$ and $W/d = 1, 2$ and $5$ (left, center, and right respectively). These correspond to data points in the $A=1$ column of \cref{fig:silo-phase-diagrams}(a).  The pressure shown here is a temporal average projected onto the mesh over several neighboring frames from the view of grid elements (from $350 \Delta t_f$ to $360 \Delta t_f$ inclusive).
    White corresponds to a completely empty element.
    (b) Plots of the pressure along the centerline for silos with $A=1.0$ and $W/d$ ratios of 1.0, 2.0 and 5.0 (pink, green, and blue respectively). While the pressure decreases drastically directly above the orifice in the clogged configurations, it does not reach zero. 
    In contrast, the flowing case has much lower pressure and is actually zero in the region above the opening.
    }
    \label{fig:silo-pressures}
\end{figure}
Next, in the top panel of \cref{fig:silo-pressures}, we observe the pressure distributions in the silos for three configurations $W/d = 1$, $2$, and $5$ with a common $A = 1$. The fields are plotted near the end of the simulation and use a spatial and temporal average over frames from $350 \Delta t_f$ to $360 \Delta t_f$. The spatial smoothing was done as in \citep{andersen09} to project point-wise pressure data onto the background mesh -- note that this is a purely visual post-processing technique and does not affect the fidelity of the simulation itself. In the jammed configurations ($W/d = 1$ and $2$), we see that the pressure immediately above the arch is low, but does not go to zero. The simulation which is closer to the static/flowing boundary ($W/d=2$) experiences a lower pressure in the region immediately above the arch, although some distance away from the opening the pressure looks essentially the same as the other static case. In contrast, the flowing simulation ($W/d = 5$) contains many materials points in the region above the orifice where material points are in free-fall, and the pressure indeed goes to zero when approaching this region since the material points are stress-free. The peak pressure is also smaller, although the fill height has already decreased substantially by the time this snapshot was taken due to the flow. The bottom panel of \cref{fig:silo-pressures} uses the same data, but shows the pressure along the vertical centerline of the silos, confirming our observations from the patch plots.
\begin{figure}
    \centering
    \begin{tabular}{c c}
       \ \ \ \ \  $A = 0.1$ & $A = 1.0$ \\
      \ \  \ \ \ \includegraphics{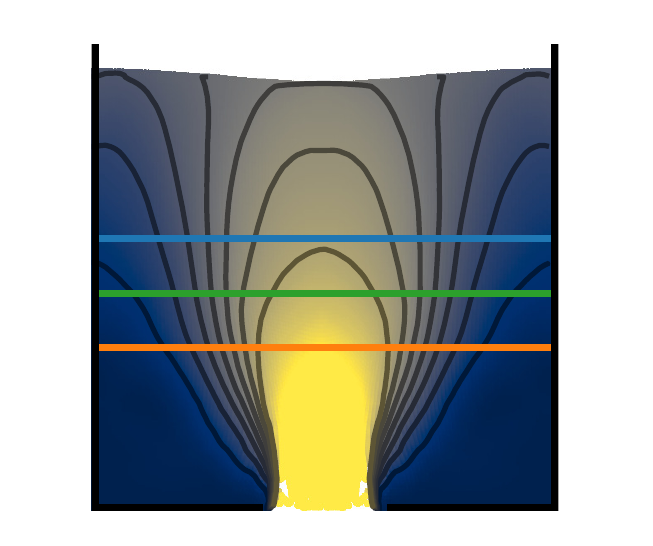} &
        \includegraphics{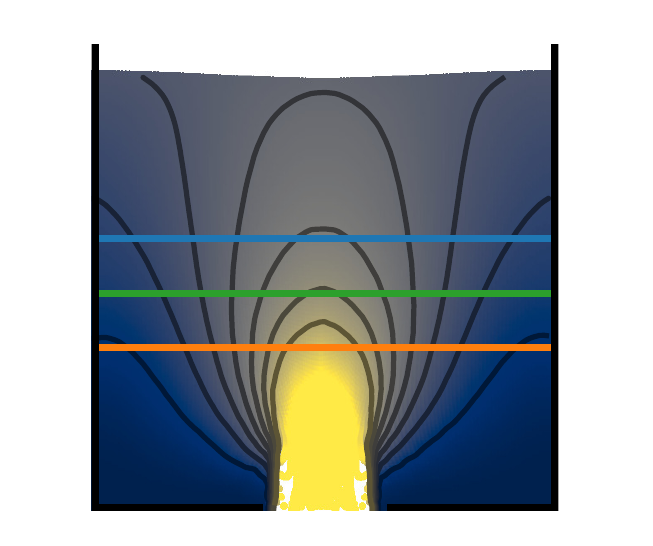} \\
    \end{tabular}
    \ \ \ \ \  \includegraphics{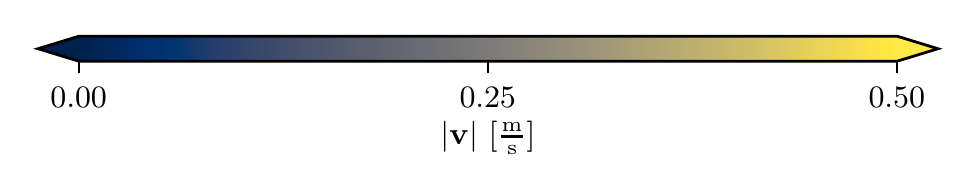}
    \begin{tabular}{c c}
        \includegraphics{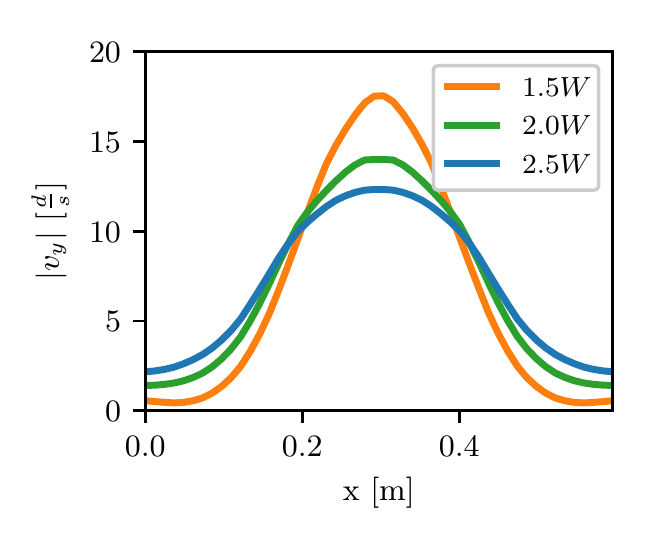} &
        \includegraphics{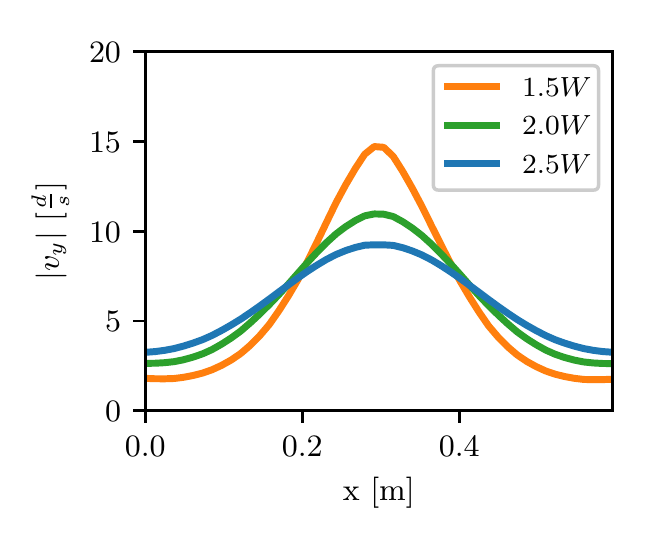}
    \end{tabular}
    \caption{Two silos with the same parameters and initial conditions except for the value of $A$. Both simulations have nearly the same amount of mass remaining in the silo when imaged ($70 \Delta t_f$ left, $86 \Delta t_f$ right). The velocity cuts off more sharply in the low $A$-value case as expected. Iso-speed lines (drawn at 0.05, 0.10, 0.15, 0.20, 0.25, 0.30, and 0.35 \si{\meter\per\second}) are drawn over a scatterplot of the speed confirm that the velocity is diffusing more when $A=1.0$ compared to when $A=0.1$. Vertical velocity profiles are drawn at the heights indicated as horizontal lines in the first row of images located at heights $1.5W, 2.0W,$ and $2.5W$ above the orifice. 
    The grain size is given by $d = \SI{0.028}{\meter}$, and the orifice size is $W = \SI{0.14}{\meter}$ resulting in $W/d = 5$.}
    \label{tab:low-A-vs-high-A-silo}
\end{figure}

Now we will turn our attention more towards the flowing simulations. In \cref{tab:low-A-vs-high-A-silo} we compare two silo flows which have the same material parameters but have different values of $A$. Snapshots are taken at slightly different times to account for the difference in outflow rates and were chosen to keep a similar amount of mass remaining in the silo; the snapshot for the $A = 0.1$ case is taken at $70 \Delta t_f$ and for the $A = 1.0$ case is taken at $86 \Delta t_f$, where again $\Delta t_f = \SI{1/240}{\second}$. In the top half of \cref{tab:low-A-vs-high-A-silo}, we see the velocity fields plotted over material points and velocity contours. The spacing of the contour lines indicates that shear zones in the $A = 0.1$ case (left) are thinner than in the $A = 1.0$ case (right). This is confirmed by plotting the vertical velocity across horizontal slices at heights $1.5W$, $2.0W$ and $2.5W$ in the lower half of the figure -- the flow is more plug-like in the low $A$ case, and is noticeably similar to a local $\mu(I)$ model solution, which would predict plug flow down the center with a thin shear band before dropping to zero vertical velocity close to the sidewalls where $\mu < \mu_s$ \citep{staron12,staron14,kamrin10}. In contrast, the higher $A$ value results in a spreading of the velocity field as expected due to the nonlocal effect, which will imply a length scale for shear bands. As both cases do actually have non-zero $A$, we do observe a small amount of flow even where $\mu < \mu_s$ as expected, and less flow under $\mu_s$ is observed for the smaller $A$. The decay length scale depends on the size of $A$, and we find that larger $A$ diffuses out the velocity more readily, also as expected.

\begin{figure}
    \centering
    \includegraphics[width=3.3in]{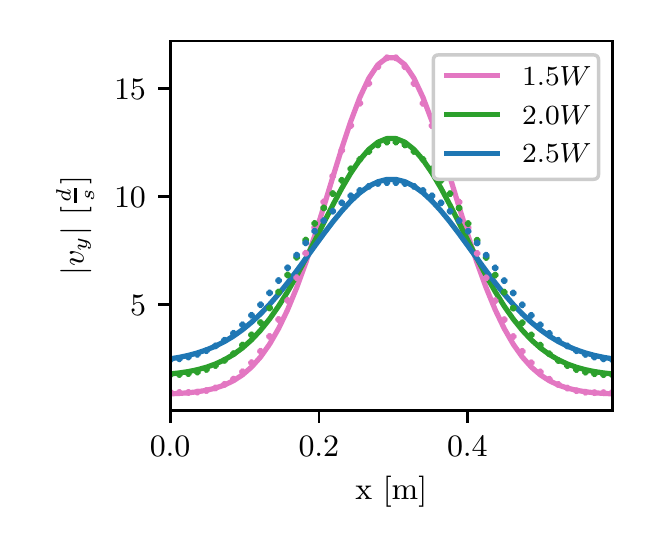}
    \caption{Cross-sections of vertical velocity taken at horizontal slices at height $1.5W$, $2.0W$, and $2.5W$ (pink, green, and blue colors respectively) in a silo with the same parameters as \cref{tab:low-A-vs-high-A-silo} except $A = 0.5$ (chosen to match experimental material). Simulation data is plotted with dots, while a diffusing Gaussian fit (with offsets) is plotted as the lines. The value of $B$, which controls the Gaussian's width at each height, is given by $0.6d$ ($d$ grain diameter). 
    }
    \label{fig:gaussian}
\end{figure}

We also look at the form of the vertical velocity profiles obtained. In order to compare to existing data, which uses glass beads primarily, we choose $A = 0.5$ and set $W/d = 5$. We waited for steady-state flow, and then looked at vertical velocity slices at the same three heights as before ($1.5W$, $2.0W$ and $2.5W$); results are plotted in \cref{fig:gaussian}. Many have noticed in experiments that the downward velocity profile, in a region not too far from the opening, appears to spread diffusively but with height playing the role of `time' --- that is, the downward velocity appears Gaussian with $v_y=C \exp{\frac{-x^2}{4By}}$ where $B$ controls the width of the Gaussian and $y$ is the height above the orifice  \citep{choi05,Rycroft2006,bazant2006spot,zuriguel14,tuzun1979experimental,samadani1999size,medina1998velocity}. The apparently diffusive character of the flow field was central to some of first silo flow models \citep{mullins1972stochastic,nedderman1979kinematic}. Interestingly, data from our simulations our also matches this observation, see \cref{fig:gaussian}, with two caveats. First, we have an additive constant at each height due to our walls having no friction (and thus allowing downward motion at the walls); in experiments, wall friction ensures velocity decays to zero at the side walls. Secondly, the width parameter that we obtain, $B = 0.6d$, is a smaller than the typical values observed in the literature ($1d-5d$ in \citet{kamrin07}). We believe this is due to the relatively large orifice size we used compared to the geometry of our system (unlike the experiments which used a much smaller opening relative to the silo width) as well as the aforementioned lack of sidewall friction.

\begin{figure}
    \centering
    \begin{tabular}{c c c}
        \includegraphics{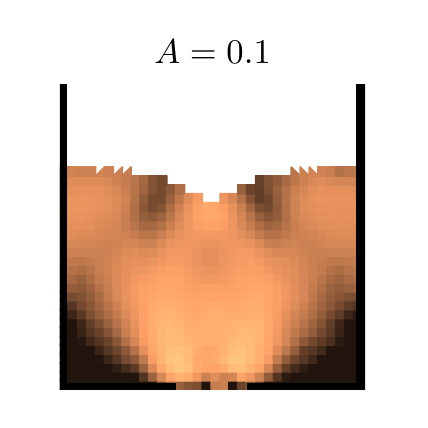} &
        \includegraphics{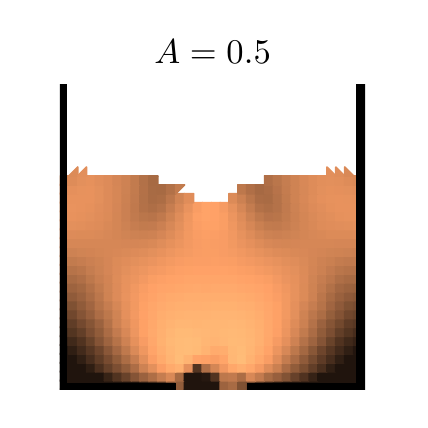} &
        \includegraphics{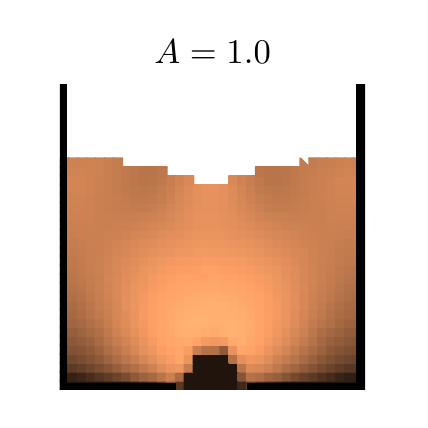} \\
    \end{tabular}
    \includegraphics{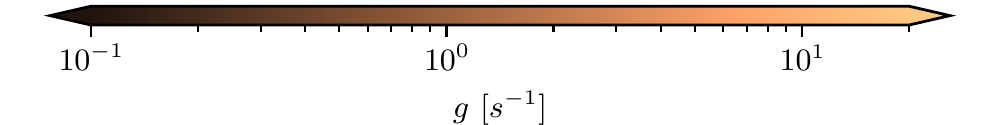}
    \caption{Plots of $g$ (logarithmic scale) of a silo with $W/d = 5$ and the same material parameters as previously used except in the value of $A$, going from $0.1$ to $0.5$ to $1.0$ from left to right. Our averaging procedure assigns the value of $g = 0$ to open material; the region immediately above the orifice appears to have low $g$ value, though it may be more correct to say there is \emph{no} value of $g$ at those points. All plots are taken over an average of frames from $280 \Delta t_f$ to $300 \Delta t_f$ and use the same mesh-wise projection technique as was used to display the pressure in \cref{fig:silo-pressures}. Note that the lower values of $A$ result in less spreading of the $g$ field as anticipated.}
    \label{fig:log-g-configuration}
\end{figure}
Finally, we plot the $g$ field in these same simulations, shown in \cref{fig:log-g-configuration}. The plots show averages over both elements and temporal frames ($280 \Delta t_f$ to $300 \Delta t_f$) to understand what the numerical scheme regularly observes when computing the spatial gradients. Note that the lower values of $A$ show less spreading of the $g$ field and a slightly larger flow-rate compared to higher values of $A$ (indicated by the height of the free surface) as expected. Considering $g$ as a surrogate for equivalent plastic shear strain rate, we can also observe the distinctive ``arms'' of high-shearing emanating from the orifice. The region just above the orifice where grains are in free-fall is also clearly visible in these plots, indicated by a dark brown color ($g$ is set to zero in the disconnected state, but those material points do not contribute to the solution of the NGF equation).

\subsection{Convergence test and caveats}\label{refine}

Convergence testing is difficult due to the amount of wall time each simulation takes and the total amount of data generated. Performing a (linearized) Von Neumann stability analysis on the scheme used for evolving the NGF PDE shows that the stable $\Delta t$ is proportional to $\Delta x^2$  due to the Laplacian term \citep{dunatunga17}. While the substepping is designed to mitigate the impact of this harsh CFL condition, running a refined simulation still takes substantially more CPU and wall time than a coarse one.

To at least show qualitative agreement, we performed another set of simulations at twice the spatial resolution of the previous cases (and correspondingly a timestep of one quarter the original size). As with the other simulations, the mesh consists of triangles stacked to create squares, and which diagonal used depends on which half of the domain the element is in. Numerical parameters are summarized in \cref{tab:refinement-parameters}. We choose $A = 1$ and other material parameters are given as before in \cref{tab:common-material-parameters}.

\begin{table}
    \centering
    \begin{tabular}{c | c | c}
         Parameter & Coarse Value & Refined Value \\
         \hline
         $\Delta x$ & \SI{0.0175}{\meter} & \SI{0.00875}{\meter} \\
         $\Delta t$ & \SI{1e-4}{\second} & \SI{2.5e-5}{\second} \\
         $t_\mathrm{end}$ & \SI{1.5}{\second} & \SI{1.5}{\second} \\
         $W$ & \SI{0.14}{\meter} & \SI{0.14}{\meter} \\
         $L$ & \SI{0.595}{\meter} & \SI{0.595}{\meter} \\
         $H$ & \SI{0.595}{\meter} & \SI{0.595}{\meter} \\
    \end{tabular}
    \caption{Numerical parameters used for the refined and coarse simulation cases used in the convergence study. The only differences are that the element size is halved and the refined timestep size is correspondingly cut down to a quarter of the coarse value. The material properties are given in \cref{tab:common-material-parameters}.}
    \label{tab:refinement-parameters}
\end{table}

Refined simulations took approximately 30 hours to complete on an Intel Silver 4112; both simulations were run at the same time, with the only difference between them being the grain diameter $d = \SI{0.014}{\meter}$ and $d = \SI{0.0028}{\meter}$ (resulting in $W/d$ = 1 and 5). See Supplemental Movie for side-by-side video of these simulations. The flowing simulation (with a smaller grain diameter) finished slightly earlier, as removed material points do not require much computational work to handle. As with the coarse silos, we used substepping with $\Delta t_{\mathrm{substep}} = \Delta t / 150$.
\begin{figure}
    \centering
    \begin{tabular}{c c}
        $W/d = 1$ & $W/d = 5$ \\
        \includegraphics{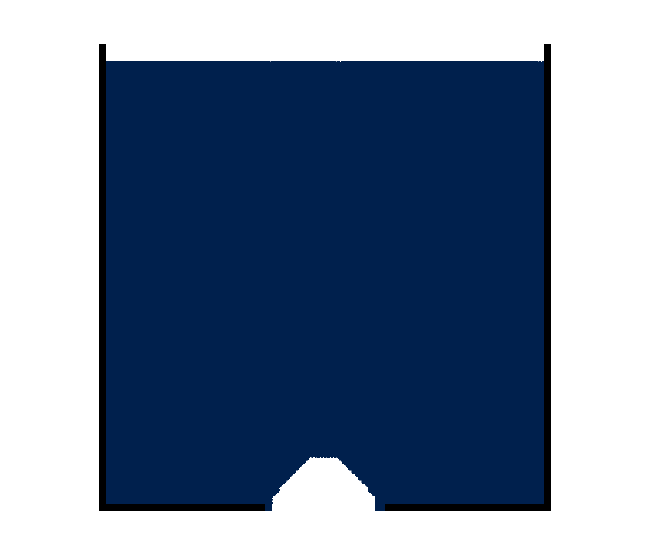} &
        \includegraphics{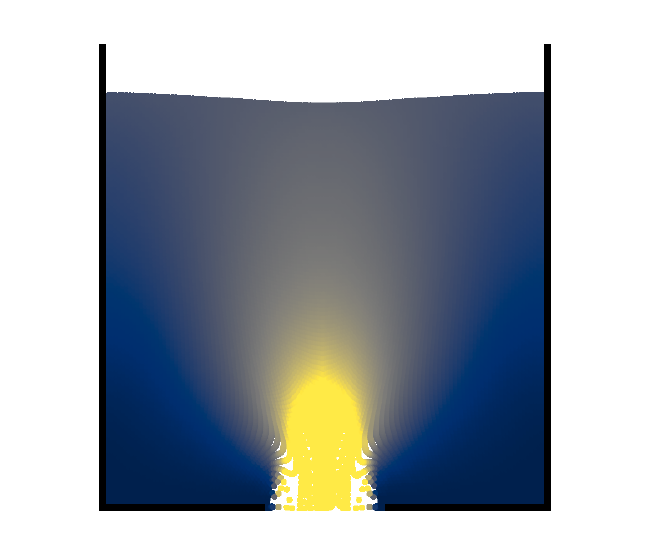} \\
        \includegraphics{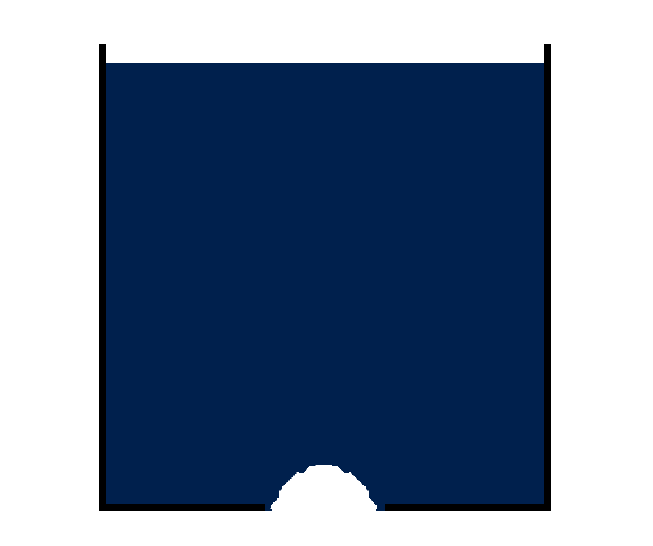} &
        \includegraphics{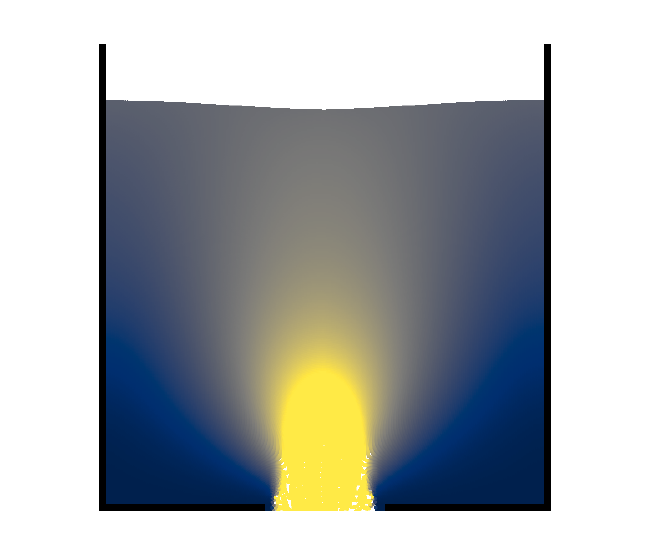}
    \end{tabular}
    \includegraphics{velocities/speed_colorbar.pdf}
    \caption{
    Velocity scatterplots of coarse simulations (top) and refined simulations (bottom). The $W/d = 1$ cases are taken at time $350 \Delta t_f$ and the $W/d = 5$ cases at $105 \Delta t_f$ (recall $\Delta t_f = \SI{1/240}{\second}$). 
    The refinement allows a more accurate representation of the orifice size across all simulations and the arch formed in the $W/d = 1$ case. 
    Velocities are plotted as given on each material point.}
    \label{fig:refined-velocity}
\end{figure}

We compared the velocity profile in this refined case to the coarse cases run previously with the results displayed in \cref{fig:refined-velocity}. Both the static configuration at $W/d = 1$ (snapshots taken at $350 \Delta t_f$, where $\Delta t_f = \SI{1/240}{\second}$ as before) and the flowing configuration at $W/d = 5$ (snapshots taken at $105 \Delta t_f$) show good agreement when refined. The increased resolution allows the arch to form with a slightly more natural circular geometry in the static case. In the flowing case, the refined simulation has a slightly higher velocity throughout the bulk, however this is expected and is also due to geometric refinement; as we measure the orifice size between the fixed points, decreasing the element size allows a larger fraction of material to leave without sensing the orifice edge. This is a minor effect, but does serve to increase the flow rate a little as observed. We expect subsequent refinements to show increasingly smaller differences in velocity fields since the affected elements will get smaller.  

\begin{figure}
    \centering
    \begin{tabular}{c c}
        $W/d = 1$ & $W/d = 5$ \\
        \includegraphics{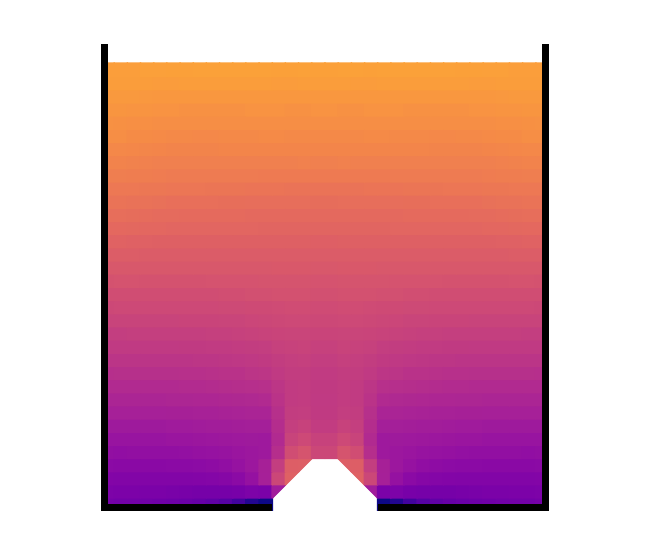} &
        \includegraphics{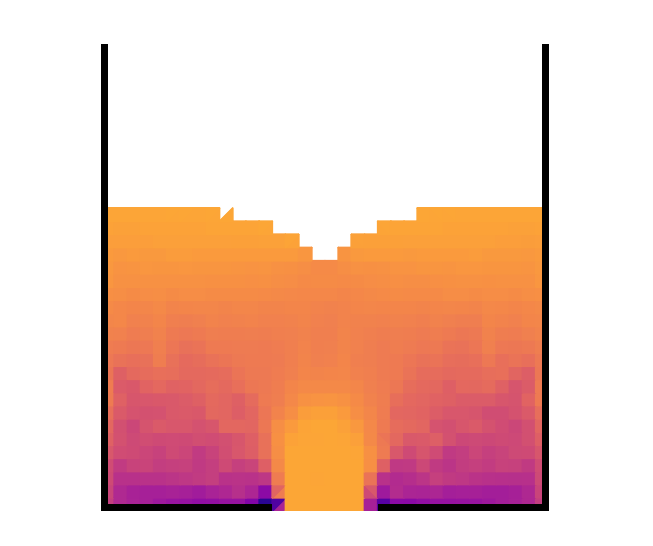} \\
        \includegraphics{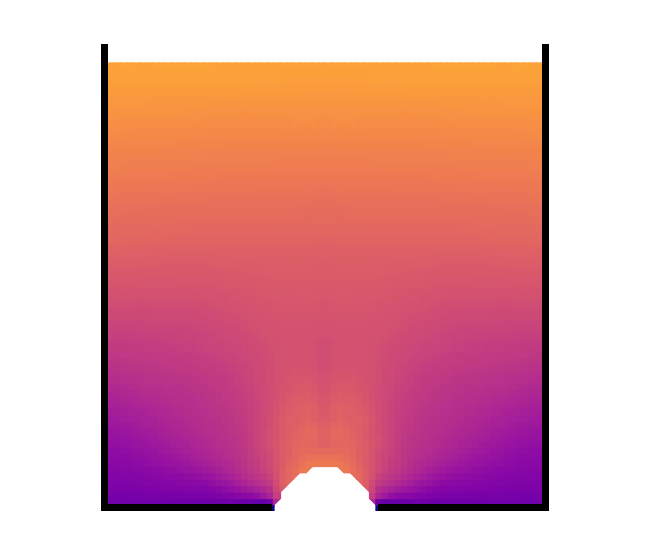} &
        \includegraphics{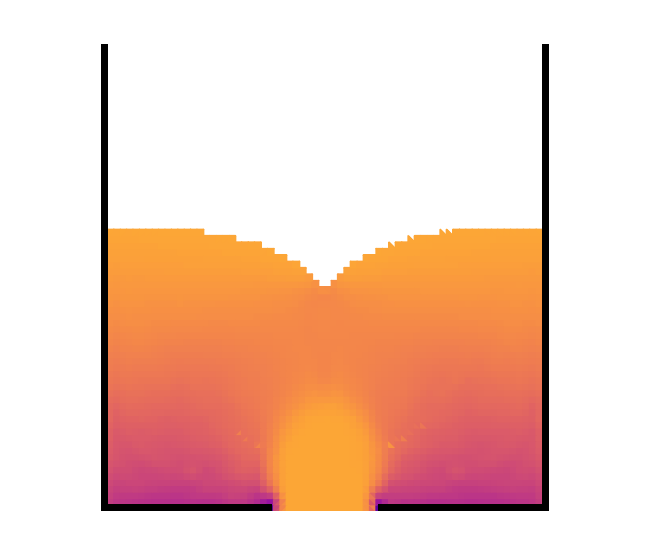}
    \end{tabular}
    \includegraphics{pressure_colorbar.pdf}
    \caption{
    Pressure distributions of coarse simulations (top) and refined simulations (bottom). All plots are taken over a three-frame average centered at $350 \Delta t_f$. 
    As with the previous pressure plots, the technique presented in \citet{andersen09} for visualization is used.
    }
    \label{fig:refined-pressure}
\end{figure}

Similarly, we compared the pressure throughout both sets of silos in \cref{fig:refined-pressure}. As before, good agreement is shown between the refined and coarse cases. The static case shows a few differences due to the more circular geometry of the arch, however the key features of small-but-nonzero pressure above the arch and a largely lithostatic pressure distribution away from the orifice remain. In the flowing case, the refined simulation has a noticeably smoother pressure field and resolves the free surface with more fidelity as expected, but otherwise qualitatively looks similar to the coarse case. We can clearly see the free-fall region above the orifice in both cases, where the pressure of the material points goes to zero as they enter the disconnected state.

Since entire elements contribute to the $g$ field and are not considered at a sub-grid scale, this in turn restricts the shapes that the boundary can take; the effect is most apparent in simulations such as a jammed silo since the arch formed must conform to the background mesh geometry. Sufficient mismatch between the true solution and the one forced by the grid geometry might result in flow where none was expected or vice-versa. We only observed this in boundary cases (e.g. in the silo, ones in which we were unsure were flowing or static), but the underlying mechanism may persist in pathological grid topologies. We also note that computing the solution is noticeably more expensive in time than a local version of the model -- the extra operations performed do contribute to this, but the increase in number of synchronization points also should not be underestimated. Naively coding the algorithm presented can easily result in a method which is tens to hundreds of times slower than the local-only, $\mu(I)$ model; while none of our implementations have been heavily optimized, many of the simulations in this work took approximately two to five times longer than our implementation in \cite{dunatunga15} of a local-only model implementing the \textbf{$\mu(I)$} rheology from \citet{jop06} (caching the results where possible during substepping results in significant time savings).

\section{Conclusion}
We have presented and numerically implemented a trans-phase (disconnection-permitting) extension of the NGF model.  Using our implementation, we have shown that NGF, a \emph{continuum} model, is able to predict the clogging of a silo due to small opening size. This result is analogous to the same model's previously shown capability to capture the strengthening of layers on inclines as they get thinner. We find that when the model's parameters are set close to the known values for spherical beads, the critical opening size obtained is close to the reported experimentally observed range and corresponds to the Beverloo cutoff size.  We have also verified by constructing a phase diagram that the criterion for silo clogging in NGF takes the form $W_{\text{crit}}\propto A d$ showing the dependence on the model's nonlocal amplitude, $A$, to be a straightforward linear prefactor, and the relevant length-scale being the mean grain size, as expected.   By including a separated phase within the rheology, the model is able to capture the process of a stable arch forming when a silo clogs, with material under the arch falling out freely. When the opening is large enough to admit a steady flow, we find the value of $A$ for a given $d$ influences how much the flow spreads spatially, with low values of $A$ emulating the purely local response with sharp flow peak in the silo center that rapidly decays toward the walls, and larger $A$ values causing a shallower peak and smoother tapering off toward the walls.  Lower $A$ values correlate with higher overall outflow rates, which is to be expected since nonlocality acts as a flow penalty for sharply varying fields.  When the parameters are set to those of spherical beads, we find, similar to previous observations for glass beads in slab silos, that the downward velocity profile along horizontal cross-sections looks similar to a Gaussian spreading diffusively in height.  

The versatility of our numerical approach, combining MPM with a simple finite-difference grid, allowed us to numerically solve the NGF model in multiple of inhomogeneous flow geometries. As a validation check, the method reproduced the analytically known velocity profiles and $H_{\text{stop}}(\theta)$ curves in the case of the inclined chute.  
Throughout our study, when comparing silos of different $W/d$, we were careful to use the same grid and initial material point distribution so as to ensure the numerical scheme does not bias the critical opening size in any suite of tests. Finally we showed that refining the silo geometry discretization results in qualitatively very similar behavior with only minor differences due to increased spatial resolution. 

This study leaves at least two interesting theoretical questions.  The first is in reference to the value of $\mu_2$ and whether this parameter, which carries over from the local $\mu(I)$ model, should in fact be bounded.  Secondly, the value of the fluidity-diffusion time-scale $t_0$ has not been previously quantified in experiments, hence our study has chosen it to be sufficiently small compared to all other time scales so as to deliver  $g$ distributions that are essentially quasi-steady-state throughout.  More study would be needed to correctly characterize this material parameter for more transient applications.\\
\\

\noindent\textbf{Declaration of Interests.} The authors report no conflict of interest. 

\appendix
\section{Discussion of alternative methods}\label{compare}
The numerical solution of the continuum equations of motion is typically done
via one of two broad classes of methods; those which track a section of the
body over time, obtaining a history of positions and velocities (Lagrangian methods), and those which track a volume of space
over time, obtaining a history of the material which entered and exited the domain (Eulerian methods).

Of the Lagrangian methods, the Finite Element Method (FEM) is widely used for
solid mechanics. 
FEM divides up a body into smaller elements, each of which can be described
with a small set of state information. FEM typically takes displacements on
nodes of these elements as the degrees of freedom, so each element tracks the
same part of the body throughout the entire simulation. This is usually
achieved by using the Galerkin Method to obtain a system of equations where the
coefficients are sums of integrals over elements (and then solving the system
via numerical linear algebraic methods). However, this places restrictions on
the types of deformations allowed. Specifically, large nonlinear deformations
are problematic, as they usually distort the mesh such that quadrature over an
element does not yield accurate results.

In contrast, Finite Volume (FV) methods are common in fluid simulations. These
are Eulerian methods in nature, performing all operations of a fixed grid.
Material may move in and out of cell boundaries, or even leave the domain of
interest entirely -- the elements are fixed in space and not tied to a
particular parcel of fluid. These methods have a complimentary set of strengths
and weaknesses as compared to FEM; they can handle large inelastic deformations
easily, but have difficulties 
representing material boundaries and material state fields, including strain, as needed for static elastic solutions.

 In most cases, the limitations of each method are less problematic than they
might first appear; simulation of solid material typically will not involve the
large nonlinear deformations that are fatal to regular FEM, and rarely does a
fluid simulation require accurate representation of a static elastic configuration. But
granular materials are an interesting exception, as they show a rich variety of
behaviors. Even within the same simulation, a granular material may flow like a
fluid, disperse into a gas-like state, and finally form a static configuration
within a short period of time -- as in the simulation of an hourglass.

 Arbitrary Lagrangian Eulerian (ALE) techniques attempt to bridge this gap. Many
of these methods are based on FEM, but with a periodic remeshing step to keep
the mesh from distorting too much. Unfortunately, this can lead to loss of
nominally conserved quantities during the remeshing process. There are also
questions as to the feasibility of material states achieved during the
remeshing process -- in general the remeshing is not aware of invariants that
must be maintained in the constitutive relations. In the context of NGF, this
may manifest as physically infeasible values for $g$ after remeshing.

\section{Integration in MPM}
\label{appendix:mpm-integration}
In weak form, balance of momentum becomes
\begin{align}
    \int_{\partial \Omega} \cauchystress \mathbf{n} \cdot \mathbf{w} \dS - \int_{\Omega} \cauchystress \colon \sGrad \mathbf{w} \dV + \int_{\Omega} \rho \mathbf{b} \cdot \mathbf{w} \dV &= \int_{\Omega} \rho \dot{\mathbf{v}} \cdot \mathbf{w} \dV,
\end{align}
where $\mathbf{w}$ is an arbitrary vector function which matches the kinematic boundary conditions.
As a special case, taking the weight function as a variation on velocity yields the principle of virtual power.
Note that this derivation is not unique to MPM; FEM uses the same weak form of momentum balance.

In FEM, these equations are discretized in space through the Galerkin method. MPM follows this formulation, but
with the key difference that numerical quadrature is done via the mobile material points whereas FEM typically fixes the quadrature points relative to the elements.

Going through the usual Galerkin procedure, we arrive at the following expression:
\begin{align}
     \sum_j \mathbf{a}_j \int_{\Omega} \rho S_j  S_i \dV &= \int_{\partial \Omega} \cauchystress \mathbf{n} S_i \dS - \int_{\Omega} \cauchystress \sGrad S_i \dV + \int_{\Omega} (\rho \mathbf{b}) S_i \dV
\end{align}
Here $S_i$ and $S_j$ represent the coefficients of shape functions on elements of a background mesh, and the equation can be compactly expressed as $\mathbf{M} \mathbf{a} = \mathbf{f}$.
 
In FEM, the integrals are typically evaluated using the shape functions via quadrature at the Gauss points, but with MPM the material points are not necessarily collocated with these Gauss points and Riemann integration is performed instead by summing over material points. While the first step of splitting the integrals element-wise is the same, yielding
\begin{align}
    \int_\Omega f(\mathbf{x}) \dV = \sum_i \int_{\Omega^e_i} f(\mathbf{x}) \dV,
\end{align}
the element integral is further split into a sum over material points as in
\begin{align}
    \int_{\Omega^e} f(\mathbf{x}) \dV \approx \sum_p \int_{\Omega^e \cap \Omega^p} f(\mathbf{x}) \dV.
\end{align}
To do this, we must consider the influence of each material point. The approximation functions have distinct error characteristics -- using a scaled delta function within a material point's region of influence recovers the original MPM formulation, while assuming a constant value over the region of integration yields methods such as generalized interpolation material point (GIMP) from \citet{bardenhagen04}, convected particle domain interpolation (CPDI) from \cite{sadeghirad11}, and descendants.

In these methods, we generally need to evaluate the shape function $S_i(\mathbf{x})$ (associated with a node of the background mesh) at positions related to $\mathbf{x}_p$ (the position of the material point). We can consider this as creating a mapping matrix $S_{ip}$ between the material points and the ephemeral mesh nodes. Similarly, we can construct gradient mappings which we write as $\nabla S_{ip}$.

Our implementation is able to switch between the various methods easily, as this typically mainly affects how the mapping matrices in MPM are constructed (i.e. the coefficients given in $S_{ip}$ and $\nabla S_{ip}$). However, for many simulations we use a variant similar to uGIMP (see \citet{sadeghirad11} for the implementation through CPDI), where each material point has an unchanging spatial extent over which the shape functions and gradient of shape functions are calculated. The underlying chunk the material point represents does allow for volume change however; in this manner, we have a tunable way to mitigate cell crossing error without incurring other types of errors induced by large inhomogeneous deformations.

Finally, note that the matrix $\mathbf{M}$, while sparse, contains off-diagonal entries and is not necessarily simple to invert. Many MPM implementations (including the present work) use an explicit forward Euler time step, so matrix solves are too expensive.
The lumped mass is typically used instead to diagonalize this matrix and allow for easy inversion:
\begin{align}
    \mathbf{M}_{\mathrm{lumped}} &= \sum_p m_p S_i(\mathbf{x}_p) \delta_{ij},
\end{align}
where $\delta_{ij}$ is the Kronecker delta (yields the value 1 when $i = j$, 0 otherwise).

Choosing when and how to update the material point state can
result in significantly different properties in the resulting numerical method.
Simulations in this work use the Update Stress Last
(USL) method coined by \citet{bardenhagen02} but first presented in \citet{sulsky94}; for comparison with other stress update schemes, refer to \citet{bardenhagen02}.
We detail our specific implementation in \cref{appendix:mpm} (except the novel constitutive update for solving the NGF model, which is presented in the main text).

\section{Details of the overall MPM procedure}
\label{appendix:mpm}

The material point method used to solve the equations of motion is fairly standard; we describe the specifics of our implementation for an MPM step below. Our novel contributions from the main text are in the solution of the constitutive update at the end of an MPM step (to implement the NGF model).

\begin{enumerate}
\item
    Compute the sparse matrices for the mapping and gradient mappings between
material points and the mesh, $\bar{S}_{ip}$ and $\sGrad \bar{S}_{ip}$.  Each
material point only contributes to a fixed, small number of nodes (those of the
element in which it is contained), so most entries in this matrix are zero
except in pathological cases.

\item
    Use the mapping matrix to compute nodal momentum and nodal mass via
    \begin{align}
        (m\mathbf{v})^n_i &= \sum_p \bar{S}_{ip} m_p \mathbf{v}^n_p \label{eqn:p1} \\
        m^n_i &= \sum_p \bar{S}_{ip} m_p.
        \label{eqn:p2}
    \end{align}

\item
    Forces on the nodes may come from roughly four sources -- material point
stresses, material point body forces, applied tractions, and contact forces.
The contact forces are important in cases such as intruder problems, but for
our purposes we only need the first three. They are computed via
    \begin{align}
        (\mathbf{f}_{\mathrm{int}})^n_i &= \sum_p \cauchystress^n_p \sGrad \bar{S}_{ip} v^n_p \\
        (\mathbf{f}_{\mathrm{ext}})^n_i &= \sum_p \bar{S}_{ip} \mathbf{b}^n_p m_p \\
        (\mathbf{f}_{\mathrm{trac}})^n_i &= \int_{\partial \Omega} \mathbf{t} S_i.
    \end{align}
The final equation is not problematic, as it consists of \emph{prescribed}
tractions, and hence the right hand side is simply a scaled integral
evaluation. If a contact algorithm were to be used, we would run it now to
compute the contact forces $(\mathbf{f}_{\mathrm{c}})^n_i$; again, this term is
zero in all simulations performed in this work.  The force contributions are
combined to obtain the force on a node of the grid
    \begin{align}
        \mathbf{f}^n_i &= (\mathbf{f}_{\mathrm{int}})^n_i + (\mathbf{f}_{\mathrm{ext}})^n_i + (\mathbf{f}_{\mathrm{trac}})^n_i + (\mathbf{f}_{\mathrm{c}})^n_i.
    \end{align}

\item
    While respecting the boundary conditions, solve the equations of motion on
the grid and deform it via
    \begin{align}
        \mathbf{a}^n_i &= \frac{\mathbf{f}^n_i}{m^n_i} \\
        (m\mathbf{v})^{n+1}_i &= (m\mathbf{v})^n_i + \Delta t \mathbf{f}^n_i \\
        \mathbf{v}^{n+1}_i &= \frac{(m\mathbf{v})^{n+1}_i}{m^n_i}.
    \end{align}

    We take the approach that when the boundary condition is fixed (Dirichlet) and
zero-displacement, both the force and momentum must be set to zero before
computing these quantities. Other implementations apply a force on the node to
cause the momentum to zero out by the end of the step. The implications of each
choice are discussed more thoroughly in \citet{buzzi08}. Natural (Neumann) and periodic conditions are implemented with ghost nodes referencing
the appropriate parts of the mesh, with a sign change if necessary.

\item
    Use the transpose mapping to update the velocity, position, and velocity
gradient on the material points via
    \begin{align}
        \mathbf{v}^{n+1}_p &= \mathbf{v}^n_p + \Delta t \sum_i \bar{S}_{ip} \mathbf{a}^n_i \\
        \mathbf{x}^{n+1}_p &= \mathbf{x}^n_p + \Delta t \sum_i \bar{S}_{ip} \mathbf{v}^{n+1}_i \\
        \velgrad^{n+1}_p &= \sum_i \mathbf{v}^{n+1}_i \otimes \sGrad \bar{S}_{ip}
    \end{align}

\item
    Use the velocity gradient on each material point to update its volume via
    \begin{align}
        v^{n+1}_p = v^n_p \exp( \Delta t \tr \velgrad^{n+1}_p ).
    \end{align}

\item
    Update the stress on each material point via
    \begin{align}
        \cauchystress^{n+1}_p = \hat{\cauchystress}(\velgrad^{n+1}_p, \dots),
    \end{align}
    where the function $\hat{\cauchystress}(\velgrad^{n+1}_p, \dots)$ is given by the constitutive relation.

\item
    Increment the time to $t = t^{n+1} = t^n + \Delta t$ and repeat until the simulation is completed.
\end{enumerate}

The above procedure can be applied with any constitutive model, although the volume may not need to be explicitly tracked for some materials; we describe the
specialization to the nonlocal granular fluidity model in the main text.

\section*{References}
\bibliographystyle{jfm}
\bibliography{references}

\begin{thebibliography}{60}
\expandafter\ifx\csname natexlab\endcsname\relax\def\natexlab#1{#1}\fi
\def\au#1{#1} \def\ed#1{#1} \def\yr#1{#1}\def\at#1{#1}\def\jt#1{\textit{#1}}
  \def\bt#1{#1}\def\bvol#1{\textbf{#1}} \def\vol#1{#1} \def\pg#1{#1}
  \def\publ#1{#1}\def\arxiv#1{#1}\def\org#1{#1}\def\st#1{\textit{#1}}

\bibitem[Abe {\em et~al.\/}(2014)Abe, Soga \& Bandara]{abe13}
{\sc \au{Abe, Keita}, \au{Soga, Kenichi} \& \au{Bandara, Samila}} \yr{2014}
  \at{Material point method for coupled hydromechanical problems}.  \jt{Journal
  of Geotechnical and Geoenvironmental Engineering}  \bvol{140}~(3),
  \pg{04013033}.

\bibitem[Andersen \& Andersen(2009)]{andersen09}
{\sc \au{Andersen, Soren} \& \au{Andersen, Lars}} \yr{2009} {Analysis of stress
  updates in the material-point method}.  \bt{In {\em Proceedings of the Twenty
  Second Nordic Seminar on Computational Mechanics\/}},  \pg{pp. 129--134}.
  Aalborg.

\bibitem[Aranson \& Tsimring(2001)]{aranson01}
{\sc \au{Aranson, Igor~S} \& \au{Tsimring, Lev~S}} \yr{2001}  \at{Continuum
  description of avalanches in granular media}.  \jt{Physical Review E}
  \bvol{64}~(2),  \pg{020301}.

\bibitem[Bandara \& Soga(2015)]{bandara15}
{\sc \au{Bandara, S.} \& \au{Soga, K.}} \yr{2015}  \at{{Coupling of soil
  deformation and pore fluid flow using Material Point Method}}.  \jt{Computers
  and Geotechnics}  \bvol{63},  \pg{199--214}.

\bibitem[Bardenhagen(2002)]{bardenhagen02}
{\sc \au{Bardenhagen, S.G.}} \yr{2002}  \at{Energy conservation error in the
  material point method for solid mechanics}.  \jt{Journal of Computational
  Physics}  \bvol{180},  \pg{383--403}.

\bibitem[Bardenhagen {\em et~al.\/}(2000)Bardenhagen, Brackbill \&
  Sulsky]{bardenhagen00}
{\sc \au{Bardenhagen, S.~G.}, \au{Brackbill, J.~U.} \& \au{Sulsky, Deborah}}
  \yr{2000}  \at{{The material-point method for granular materials}}.
  \jt{Computer Methods in Applied Mechanics and Engineering}  \bvol{187}~(3-4),
   \pg{529--541}.

\bibitem[Bardenhagen \& Kober(2004)]{bardenhagen04}
{\sc \au{Bardenhagen, S.~G.} \& \au{Kober, E.~M.}} \yr{2004}  \at{{The
  generalized interpolation material point method}}.  \jt{Computer Modeling in
  Engineering \& Sciences}  \bvol{5}~(6),  \pg{477--495}.

\bibitem[Bazant(2006)]{bazant2006spot}
{\sc \au{Bazant, Martin~Z}} \yr{2006}  \at{The spot model for random-packing
  dynamics}.  \jt{Mechanics of Materials}  \bvol{38}~(8-10),  \pg{717--731}.

\bibitem[Beverloo {\em et~al.\/}(1961)Beverloo, Leniger \& van~de
  Velde]{beverloo61}
{\sc \au{Beverloo, W.~A.}, \au{Leniger, H.~A.} \& \au{van~de Velde, J.}}
  \yr{1961}  \at{{The flow of granular solids through orifices}}.  \jt{Chemical
  Engineering Science}  \bvol{15}~(3-4),  \pg{260--269}.

\bibitem[Buzzi {\em et~al.\/}(2008)Buzzi, Pedroso \& Giacomini]{buzzi08}
{\sc \au{Buzzi, Olivier}, \au{Pedroso, Dorival~M} \& \au{Giacomini, Anna}}
  \yr{2008}  \at{{Caveats on the implementation of the generalized material
  point method}}.  \jt{Computer Modeling in Engineering \& Sciences}
  \bvol{1}~(1),  \pg{1--21}.

\bibitem[Choi {\em et~al.\/}(2005)Choi, Kudrolli \& Bazant]{choi05}
{\sc \au{Choi, Jaehyuk}, \au{Kudrolli, Arshad} \& \au{Bazant, Martin~Z}}
  \yr{2005}  \at{{Velocity profile of granular flows inside silos and
  hoppers}}.  \jt{Journal of Physics: Condensed Matter}  \bvol{17}~(24),
  \pg{S2533--S2548}.

\bibitem[{da Cruz} {\em et~al.\/}(2005){da Cruz}, Emam, Prochnow, Roux \&
  Chevoir]{dacruz05}
{\sc \au{{da Cruz}, Fr\'{e}d\'{e}ric}, \au{Emam, Sacha}, \au{Prochnow,
  Micha\"{e}l}, \au{Roux, Jean-No\"{e}l} \& \au{Chevoir, Fran\c{c}ois}}
  \yr{2005}  \at{{Rheophysics of dense granular materials: Discrete simulation
  of plane shear flows}}.  \jt{Physical Review E}  \bvol{72}~(2),  \pg{021309}.

\bibitem[Dunatunga \& Kamrin(2015)]{dunatunga15}
{\sc \au{Dunatunga, Sachith} \& \au{Kamrin, Ken}} \yr{2015}  \at{Continuum
  modelling and simulation of granular flows through their many phases}.
  \jt{Journal of Fluid Mechanics}  \bvol{779},  \pg{483--513}.

\bibitem[Dunatunga \& Kamrin(2017)]{dunatunga16}
{\sc \au{Dunatunga, Sachith} \& \au{Kamrin, Ken}} \yr{2017}  \at{Continuum
  modeling of projectile impact and penetration in dry granular media}.
  \jt{Journal of the Mechanics and Physics of Solids}  \bvol{100},  \pg{45 --
  60}.

\bibitem[Dunatunga(2017)]{dunatunga17}
{\sc \au{Dunatunga, Sachith~Anurudde}} \yr{2017}  \at{A framework for continuum
  simulation of granular flow}. PhD thesis, Massachusetts Institute of
  Technology.

\bibitem[de~Gennes(1999)]{degennes99}
{\sc \au{de~Gennes, P.}} \yr{1999}  \at{{Granular matter: a tentative view}}.
  \jt{Reviews of Modern Physics}  \bvol{71}~(2),  \pg{S374--S382}.

\bibitem[Gurtin {\em et~al.\/}(2010)Gurtin, Fried \& Anand]{gurtin10}
{\sc \au{Gurtin, Morton~E.}, \au{Fried, Eliot} \& \au{Anand, Lallit}} \yr{2010}
  {\em {The mechanics and thermodynamics of continua}\/}.  \publ{Cambridge
  University Press}.

\bibitem[Henann \& Kamrin(2013)]{henann13}
{\sc \au{Henann, David~L} \& \au{Kamrin, Ken}} \yr{2013}  \at{{A predictive,
  size-dependent continuum model for dense granular flows.}}  \jt{Proceedings
  of the National Academy of Sciences of the United States of America}
  \bvol{110}~(17),  \pg{6730--6735}.

\bibitem[Henann \& Kamrin(2014)]{henann14}
{\sc \au{Henann, David~L.} \& \au{Kamrin, Ken}} \yr{2014}  \at{{Continuum
  thermomechanics of the nonlocal granular rheology}}.  \jt{International
  Journal of Plasticity}  \bvol{60},  \pg{145--162}.

\bibitem[Hidalgo {\em et~al.\/}(2013)Hidalgo, Lozano, Zuriguel \&
  Garcimart{\'\i}n]{hidalgo13}
{\sc \au{Hidalgo, RC}, \au{Lozano, C}, \au{Zuriguel, I} \&
  \au{Garcimart{\'\i}n, A}} \yr{2013}  \at{Force analysis of clogging arches in
  a silo}.  \jt{Granular Matter}  \bvol{15}~(6),  \pg{841--848}.

\bibitem[Holyoake \& McElwaine(2012)]{holyoake12}
{\sc \au{Holyoake, Alex~J} \& \au{McElwaine, Jim~N}} \yr{2012}  \at{High-speed
  granular chute flows}.  \jt{Journal of Fluid Mechanics}  \bvol{710},
  \pg{35--71}.

\bibitem[Jop {\em et~al.\/}(2005)Jop, Forterre \& Pouliquen]{jop05}
{\sc \au{Jop, Pierre}, \au{Forterre, Yo\"{e}l} \& \au{Pouliquen, Olivier}}
  \yr{2005}  \at{{Crucial role of sidewalls in granular surface flows:
  consequences for the rheology}}.  \jt{Journal of Fluid Mechanics}
  \bvol{541},  \pg{167}.

\bibitem[Jop {\em et~al.\/}(2006)Jop, Forterre \& Pouliquen]{jop06}
{\sc \au{Jop, Pierre}, \au{Forterre, Yo\"{e}l} \& \au{Pouliquen, Olivier}}
  \yr{2006}  \at{{A constitutive law for dense granular flows.}}  \jt{Nature}
  \bvol{441}~(7094),  \pg{727--30}.

\bibitem[Kamrin(2010)]{kamrin10}
{\sc \au{Kamrin, Ken}} \yr{2010}  \at{{Nonlinear elasto-plastic model for dense
  granular flow}}.  \jt{International Journal of Plasticity}  \bvol{26}~(2),
  \pg{167--188}.

\bibitem[Kamrin(2019)]{kamrin2019non}
{\sc \au{Kamrin, Ken}} \yr{2019}  \at{Non-locality in granular flow:
  Phenomenology and modeling approaches}.  \jt{Frontiers in Physics}  \bvol{7},
   \pg{116}.

\bibitem[Kamrin(2020)]{kamrin2020quantitative}
{\sc \au{Kamrin, Ken}} \yr{2020}  \at{Quantitative rheological model for
  granular materials: The importance of particle size}.  \jt{Handbook of
  Materials Modeling: Applications: Current and Emerging Materials}  \pg{pp.
  153--176}.

\bibitem[Kamrin \& Bazant(2007)]{kamrin07}
{\sc \au{Kamrin, Ken} \& \au{Bazant, Martin~Z.}} \yr{2007}  \at{{Stochastic
  flow rule for granular materials}}.  \jt{Physical Review E}  \bvol{75}~(4),
  \pg{041301}.

\bibitem[Kamrin \& Henann(2015)]{kamrin15}
{\sc \au{Kamrin, Ken} \& \au{Henann, David~L.}} \yr{2015}  \at{Nonlocal
  modeling of granular flows down inclines}.  \jt{Soft Matter}  \bvol{11},
  \pg{179--185}.

\bibitem[Kamrin \& Koval(2012)]{kamrin12}
{\sc \au{Kamrin, Ken} \& \au{Koval, Georg}} \yr{2012}  \at{{Nonlocal
  Constitutive Relation for Steady Granular Flow}}.  \jt{Physical Review
  Letters}  \bvol{108}~(17),  \pg{178301+}.

\bibitem[Kamrin \& Koval(2014)]{kamrin14surffric}
{\sc \au{Kamrin, Ken} \& \au{Koval, Georg}} \yr{2014}  \at{Effect of particle
  surface friction on nonlocal constitutive behavior of flowing granular
  media}.  \jt{Computational Particle Mechanics}  \bvol{1}~(2),  \pg{169--176}.

\bibitem[Liu \& Henann(2018)]{liu2018size}
{\sc \au{Liu, Daren} \& \au{Henann, David~L}} \yr{2018}  \at{Size-dependence of
  the flow threshold in dense granular materials}.  \jt{Soft matter}
  \bvol{14}~(25),  \pg{5294--5305}.

\bibitem[Mankoc {\em et~al.\/}(2007)Mankoc, Janda, Arevalo, Pastor, Zuriguel,
  Garcimart{\'\i}n \& Maza]{Mankoc2007}
{\sc \au{Mankoc, C}, \au{Janda, A}, \au{Arevalo, Roberto}, \au{Pastor, JM},
  \au{Zuriguel, Iker}, \au{Garcimart{\'\i}n, A} \& \au{Maza, Diego}} \yr{2007}
  \at{The flow rate of granular materials through an orifice}.  \jt{Granular
  Matter}  \bvol{9}~(6),  \pg{407--414}.

\bibitem[Martin {\em et~al.\/}(2009)Martin, Dubois, Monerie \&
  Radjai]{martin2009jamming}
{\sc \au{Martin, A.}, \au{Dubois, F.}, \au{Monerie, Y.} \& \au{Radjai, F.}}
  \yr{2009}  \at{Jamming and flow statistics in a silo geometry}.  \jt{AIP
  Conference Proceedings}  \bvol{1145}~(1),  \pg{653--656},  \arxiv{arXiv:
  https://aip.scitation.org/doi/pdf/10.1063/1.3180011}.

\bibitem[Mast(2013)]{Mast2013}
{\sc \au{Mast, Carter~M}} \yr{2013}  \at{{Modeling Landslide-Induced Flow
  Interactions with Structures using the Material Point Method}}. PhD thesis,
  University of Washington.

\bibitem[Mast {\em et~al.\/}(2015)Mast, Arduino, Mackenzie-Helnwein \&
  Miller]{mast15}
{\sc \au{Mast, Carter~M.}, \au{Arduino, Pedro}, \au{Mackenzie-Helnwein, Peter}
  \& \au{Miller, Gregory~R.}} \yr{2015}  \at{{Simulating granular column
  collapse using the Material Point Method}}.  \jt{Acta Geotechnica}
  \bvol{10},  \pg{101--116}.

\bibitem[Medina {\em et~al.\/}(1998)Medina, Cordova, Luna \&
  Trevino]{medina1998velocity}
{\sc \au{Medina, A}, \au{Cordova, JA}, \au{Luna, E} \& \au{Trevino, C}}
  \yr{1998}  \at{Velocity field measurements in granular gravity flow in a near
  2d silo}.  \jt{Physics Letters A}  \bvol{250}~(1-3),  \pg{111--116}.

\bibitem[MiDi(2004)]{midi04}
{\sc \au{MiDi, G. D.~R.}} \yr{2004}  \at{{On dense granular flows.}}  \jt{The
  European physical journal. E, Soft matter}  \bvol{14}~(4),  \pg{341--65}.

\bibitem[Mullins(1972)]{mullins1972stochastic}
{\sc \au{Mullins, WW}} \yr{1972}  \at{Stochastic theory of particle flow under
  gravity}.  \jt{Journal of Applied Physics}  \bvol{43}~(2),  \pg{665--678}.

\bibitem[Nedderman \& T{\"u}z{\"u}n(1979)]{nedderman1979kinematic}
{\sc \au{Nedderman, RM} \& \au{T{\"u}z{\"u}n, U}} \yr{1979}  \at{A kinematic
  model for the flow of granular materials}.  \jt{Powder Technology}
  \bvol{22}~(2),  \pg{243--253}.

\bibitem[Pouliquen(1999)]{pouliquen99}
{\sc \au{Pouliquen, Olivier}} \yr{1999}  \at{Scaling laws in granular flows
  down rough inclined planes}.  \jt{Physics of Fluids (1994-present)}
  \bvol{11}~(3),  \pg{542--548}.

\bibitem[Rycroft {\em et~al.\/}(2006)Rycroft, Bazant, Grest \&
  Landry]{Rycroft2006}
{\sc \au{Rycroft, Chris~H.}, \au{Bazant, Martin~Z.}, \au{Grest, Gary~S.} \&
  \au{Landry, James~W.}} \yr{2006}  \at{{Dynamics of random packings in
  granular flow.}}  \jt{Physical review. E, Statistical, nonlinear, and soft
  matter physics}  \bvol{73}~(5 Pt 1),  \pg{051306}.

\bibitem[Sadeghirad {\em et~al.\/}(2011)Sadeghirad, Brannon \&
  Burghardt]{sadeghirad11}
{\sc \au{Sadeghirad, A.}, \au{Brannon, Rebecca~M.} \& \au{Burghardt, J.}}
  \yr{2011}  \at{{A convected particle domain interpolation technique to extend
  applicability of the material point method for problems involving massive
  deformations}}.  \jt{International Journal for Numerical Methods in
  Engineering}  \bvol{86}~(12),  \pg{1435--1456}.

\bibitem[Samadani {\em et~al.\/}(1999)Samadani, Pradhan \&
  Kudrolli]{samadani1999size}
{\sc \au{Samadani, Azadeh}, \au{Pradhan, A} \& \au{Kudrolli, A}} \yr{1999}
  \at{Size segregation of granular matter in silo discharges}.  \jt{Physical
  Review E}  \bvol{60}~(6),  \pg{7203}.

\bibitem[Sheldon \& Durian(2010)]{sheldon10}
{\sc \au{Sheldon, Hannah~G} \& \au{Durian, Douglas~J}} \yr{2010}  \at{Granular
  discharge and clogging for tilted hoppers}.  \jt{Granular Matter}
  \bvol{12}~(6),  \pg{579--585}.

\bibitem[Silbert {\em et~al.\/}(2003)Silbert, Landry \& Grest]{silbert03}
{\sc \au{Silbert, Leonardo~E}, \au{Landry, James~W} \& \au{Grest, Gary~S}}
  \yr{2003}  \at{Granular flow down a rough inclined plane: transition between
  thin and thick piles}.  \jt{Phys. Fluids}  \bvol{15}~(1),  \pg{1--10}.

\bibitem[Staron {\em et~al.\/}(2012)Staron, Lagr\'{e}e \& Popinet]{staron12}
{\sc \au{Staron, L.}, \au{Lagr\'{e}e, P.-Y.} \& \au{Popinet, S.}} \yr{2012}
  \at{{The granular silo as a continuum plastic flow: The hour-glass vs the
  clepsydra}}.  \jt{Physics of Fluids}  \bvol{24}~(10),  \pg{103301}.

\bibitem[Staron {\em et~al.\/}(2014)Staron, Lagr\'{e}e \& Popinet]{staron14}
{\sc \au{Staron, L}, \au{Lagr\'{e}e, P-Y} \& \au{Popinet, S}} \yr{2014}
  \at{{Continuum simulation of the discharge of the granular silo: a validation
  test for the $\mu$(I) visco-plastic flow law.}}  \jt{The European physical
  journal. E, Soft matter}  \bvol{37}~(1),  \pg{5}.

\bibitem[Sulsky {\em et~al.\/}(1994)Sulsky, Chen \& Schreyer]{sulsky94}
{\sc \au{Sulsky, Deborah}, \au{Chen, Zhen} \& \au{Schreyer, Howard~L.}}
  \yr{1994}  \at{{A particle method for history-dependent materials}}.
  \jt{Computer Methods in Applied Mechanics and Engineering}  \bvol{118}~(1-2),
   \pg{179--196}.

\bibitem[Thomas \& Durian(2013)]{thomas13}
{\sc \au{Thomas, Charles} \& \au{Durian, Douglas}} \yr{2013}  \at{Geometry
  dependence of the clogging transition in tilted hoppers}.  \jt{Physical
  review. E, Statistical, nonlinear, and soft matter physics}  \bvol{87},
  \pg{052201}.

\bibitem[Thomas \& Durian(2016)]{thomas16}
{\sc \au{Thomas, Charles} \& \au{Durian, Douglas}} \yr{2016}  \at{Intermittency
  and velocity fluctuations in hopper flows prone to clogging}.  \jt{Physical
  Review E}  \bvol{94}.

\bibitem[To {\em et~al.\/}(2001)To, Lai \& Pak]{to01}
{\sc \au{To, Kiwing}, \au{Lai, Pik-Yin} \& \au{Pak, H.~K.}} \yr{2001}
  \at{Jamming of granular flow in a two-dimensional hopper}.  \jt{Phys. Rev.
  Lett.}  \bvol{86},  \pg{71--74}.

\bibitem[T{\"u}z{\"u}n \& Nedderman(1979)]{tuzun1979experimental}
{\sc \au{T{\"u}z{\"u}n, U} \& \au{Nedderman, RM}} \yr{1979}  \at{Experimental
  evidence supporting kinematic modelling of the flow of granular media in the
  absence of air drag}.  \jt{Powder Technology}  \bvol{24}~(2),  \pg{257--266}.

\bibitem[Więckowski(1999)]{wieckowski99}
{\sc \au{Więckowski, Zdzisław}} \yr{1999}  \at{{A particle-in-cell solution
  to the silo discharging problem}}.  \jt{International Journal for Numerical
  Methods in Engineering}  \bvol{1225}~(February 1998),  \pg{1203--1225}.

\bibitem[Więckowski(2001)]{Wieckowski2001}
{\sc \au{Więckowski, Zdzisław}} \yr{2001} {Analysis of granular flow by the
  material point method}.  \bt{In {\em European Conference on Computational
  Mechanics\/}}.

\bibitem[Więckowski(2003)]{wieckowski03}
{\sc \au{Więckowski, Zdzisław}} \yr{2003}  \at{{Modelling of silo discharge
  and filling problems by the material point method}}.  \jt{Task Quarterly}
  \bvol{4}~(4),  \pg{701--721}.

\bibitem[Więckowski \& Kowalska-Kubsik(2011)]{wieckowski11}
{\sc \au{Więckowski, Zdzisław} \& \au{Kowalska-Kubsik, Iwona}} \yr{2011}
  {Non-local approach in modelling of granular flow by the material point
  method}.  \bt{In {\em Computer Methods in Mechanics\/}}. Warsaw, Poland.

\bibitem[Yue {\em et~al.\/}(2018)Yue, Smith, Chen, Chantharayukhonthorn, Kamrin
  \& Grinspun]{yue06}
{\sc \au{Yue, Yonghao}, \au{Smith, Breannan}, \au{Chen, Peter~Yichen},
  \au{Chantharayukhonthorn, Maytee}, \au{Kamrin, Ken} \& \au{Grinspun, Eitan}}
  \yr{2018}  \at{Hybrid grains: Adaptive coupling of discrete and continuum
  simulations of granular media}.  \jt{ACM Trans. Graph.}  \bvol{37}~(6),
  \pg{283:1--283:19}.

\bibitem[Zuriguel {\em et~al.\/}(2005)Zuriguel, Garcimart\'{\i}n, Maza,
  Pugnaloni \& Pastor]{zuriguel05}
{\sc \au{Zuriguel, Iker}, \au{Garcimart\'{\i}n, Angel}, \au{Maza, Diego},
  \au{Pugnaloni, Luis~A.} \& \au{Pastor, J.~M.}} \yr{2005}  \at{Jamming during
  the discharge of granular matter from a silo}.  \jt{Phys. Rev. E}  \bvol{71},
   \pg{051303}.

\bibitem[Zuriguel {\em et~al.\/}(2014)Zuriguel, Parisi, Hidalgo, Lozano, Janda,
  Gago, Peralta, Ferrer, Pugnaloni, Cl{\'e}ment {\em et~al.\/}]{zuriguel14}
{\sc \au{Zuriguel, Iker}, \au{Parisi, Daniel~Ricardo}, \au{Hidalgo,
  Ra{\'u}l~Cruz}, \au{Lozano, Celia}, \au{Janda, Alvaro}, \au{Gago,
  Paula~Alejandra}, \au{Peralta, Juan~Pablo}, \au{Ferrer, Luis~Miguel},
  \au{Pugnaloni, Luis~Ariel}, \au{Cl{\'e}ment, Eric} \& \au{others}} \yr{2014}
  \at{Clogging transition of many-particle systems flowing through
  bottlenecks}.  \jt{Scientific reports}  \bvol{4}~(1),  \pg{1--8}.

\bibitem[Zuriguel {\em et~al.\/}(2003)Zuriguel, Pugnaloni, Garcimartín \&
  Maza]{zuriguel03}
{\sc \au{Zuriguel, Iker}, \au{Pugnaloni, Luis}, \au{Garcimartín, Angel} \&
  \au{Maza, Diego}} \yr{2003}  \at{Jamming during the discharge of grains from
  a silo described as a percolating transition}.  \jt{Physical review. E,
  Statistical, nonlinear, and soft matter physics}  \bvol{68},  \pg{030301}.

\end{thebibliography}

\end{document}